\documentclass[prb,nobibnotes,superscriptaddress,
twocolumn,longbibliography,amsmath,amsmath,amssymb]{revtex4-1}

\usepackage[utf8]{inputenc}
\usepackage[english]{babel}

\usepackage{bm}

\usepackage{graphicx}

\usepackage[hyperindex,colorlinks,hyperfootnotes = false,citecolor = blue,]{hyperref}
\usepackage[usenames,dvipsnames,svgnames,table]{xcolor}

\newcommand{\vABnoise}{\langle v_\text{AB}(S_\eta) \rangle_\xi}
\newcommand{\vAB}{v_\text{AB}}
\newcommand{\vKS}{v_\text{KS}}
\newcommand{\vKSopt}{v_\text{KS}^{\,\ast} }
\newcommand{\Vmu}{V_\mu }
\newcommand{\Vmuopt}{V_\mu^{\,\ast} }
\newcommand{\vcorr}{\Delta v_\mu}

\usepackage{natbib}

\begin{document}

\nocite{*}

\title{
Cellular memory enhances bacterial chemotactic navigation in rugged environments
}

\author{Adam Gosztolai}
\email{adam.gosztolai@epfl.ch, Current address: Brain Mind Institute and Interfaculty Institute of Bioengineering, \'{E}cole Polyt\'{e}chnique F\'{e}d\'{e}rale de Lausanne, Lausanne, Switzerland}
\affiliation{Department of Mathematics, Imperial College London, SW7 2AZ, London, United Kingdom}
%
\author{Mauricio Barahona}
\email{m.barahona@imperial.ac.uk}
\affiliation{Department of Mathematics, Imperial College London, SW7 2AZ, London, United Kingdom}

\date{\today}

\begin{abstract}
The response of microbes to external signals is mediated by biochemical networks with intrinsic time scales. These time scales give rise to a memory that impacts cellular behaviour. Here we study theoretically the role of cellular memory in \textit{Escherichia coli} chemotaxis. Using an agent-based model, we show that cells with memory navigating rugged chemoattractant landscapes can enhance their drift speed by extracting information from environmental correlations.  Maximal advantage is achieved when the memory is comparable to the time scale of fluctuations as perceived during swimming. We derive an analytical approximation for the drift velocity in rugged landscapes that explains the enhanced velocity, and recovers standard Keller-Segel gradient-sensing results in the limits when memory and fluctuation time scales are well separated. Our numerics also show that cellular memory can induce bet-hedging at the population level resulting in long-lived multi-modal distributions in heterogeneous landscapes.
\end{abstract}

\maketitle
    
\section*{Introduction}    
    
The natural habitat of many microbes is shaped by inherent micro-scale ruggedness arising from random spatial inhomogeneities due to porous or particulate matter~\cite{Blackburn,Ngom,Hol}) or to filamentous structures resulting from turbulent advection in the medium~\cite{torney,stocker}. Microbes typically navigate such rugged attractant landscapes in search of nutrients and stimulants in a process called chemotaxis. Chemotaxis is mediated and governed by specialised biochemical pathways that sense changes in stimulant concentration, transduce those signals, and induce subsequent adjustments to the locomotion of the cell~\cite{Porter2011}. Such pathways have characteristic dynamic responses with intrinsic time scales, which are used by cells to resolve changes in chemoattractant concentrations, i.e., to perform local gradient-sensing~\cite{macnab,berg:93}. In addition, the dynamic response of the biochemical circuits can filter out the high frequencies of noisy signals, to enhance gradient-sensing~\cite{block,andrews,aquino}. 

The time scales of such responses can also be viewed as the basis for a cellular memory, over which signals are processed. Indeed, microbes sample continuously their chemical environment along their swimming trajectory, and recent work has shown that the biochemical memory can be dynamically tuned~\cite{Shah} from seconds to minutes~\cite{sourjik} in response to environmental statistics. Hence, in addition to evaluating the stimulant gradient, cells could extract informative features of the heterogeneous environment from the fluctuations they perceive as they swim.  

We study the effects of cellular memory in the context of \textit{Escherichia coli} (\textit{E. coli}) chemotaxis, a model system for the navigation of microbes~\cite{stocker}, worms~\cite{Pierce}, and eukaryotes~\cite{polin}, as well as an inspiration for the motion of swarm robots~\cite{Taylor-King,ebbens} and random search algorithms~\cite{viswanathan}. \textit{E. coli} chemotaxis entails a run-and-tumble strategy: runs (i.e., stretches of linear motion at constant velocity) interrupted by tumbles (i.e., random stops with reorientation onto a random direction). To generate a drift towards high chemoattractant concentrations, cells reduce their tumbling rate upon sensing a favourable gradient, thus lengthening the up-gradient runs~\cite{berg:72}.

The tumble rate is regulated by a chemotactic pathway with a bi-lobed temporal response with a characteristic time scale $\gamma$, which we denote the cellular memory.  Input signals are convolved with this temporal response, with the effect that recent samples are weighed positively whereas signals in the past are given a negative weighting~\cite{berg:86}. It has been shown that this response yields an estimate of the local temporal gradient~\cite{block,berg:93}. 

The capability of cells to compute local gradients is the basis for several coarse-grained models (drift-diffusion equations). The classic example is the linear Keller-Segel (KS) model~\cite{KS,othmer2}, which describes the behaviour of a population of cells whose mean velocity aligns instantaneously with the local gradient. The KS model successfully reproduces a variety of chemotactic phenomena, including experimentally observed distributions under shallow attractant gradients~\cite{menolascina}. Yet the presence of fluctuations  may lead to an incorrect assessment of the underlying gradient if using only instantaneous information. Indeed, KS fails to recapitulate situations when cells do not have time to adapt to large fluctuations, both in experiments~\cite{tu2} and in agent-based simulations~\cite{xue,dufour,long}. These shortcomings suggest the need to consider additional time scales  that play a role in chemotactic transient responses~\cite{ONpaper}, and, specifically, the intrinsic memory of the chemotactic pathway processing incoming stimuli~\cite{tu,lambert}. 

Here, we study how bacteria use their cellular memory as they swim across a rugged chemoattractant landscape to extract spatio-temporal information from the perceived signal to improve their chemotactic navigation. To shed light on the role of memory, we carry out simulations of an agent-based (AB) model containing an input-output response function of the \textit{E. coli} chemotactic pathway~\cite{sontag,becker,jiang} and compare its predictions to the KS model, which is based on memoryless local gradient alignment. The KS agrees well with the AB numerics for constant gradients, yet it underestimates the drift velocity of the population when the ambient concentration has spatial correlations, consistent with cells taking advantage of correlations in addition to local gradients.  
	
Motivated by these numerical findings, we derive an analytical formula for the drift velocity in terms of the cellular memory and the length scale of the spatial correlations of the attractant landscape. Our model predicts the numerical results and recovers KS in various limits, thus elucidating the conditions in which cellular memory provides a chemotactic advantage over memoryless local gradient-sensing. We also show that our results are consistent with optimal information coding by the chemotaxis pathway~\cite{endres,becker}, yet cells are band-limited by their tumbling rate. Our work thus extends the gradient-sensing viewpoint in chemotaxis, and provides insight into the role of memory in navigating heterogeneous landscapes. 

\section*{Results}

\section*{Gradient-sensing as the classical viewpoint of chemotaxis}

\begin{figure}[t!]
\centering
\includegraphics[width=.95\columnwidth]{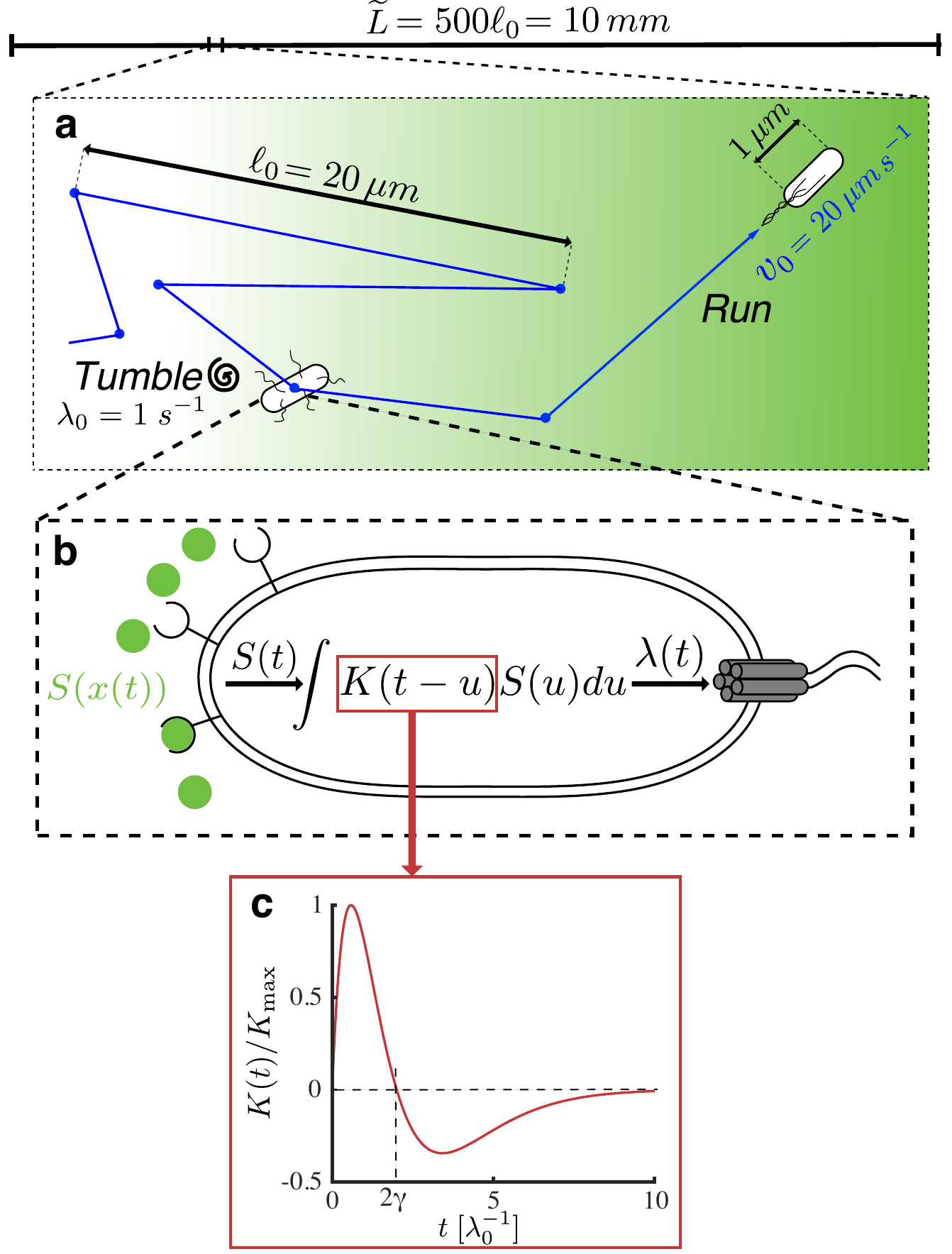}
\caption{\textbf{Setup of the agent-based model and simulation framework.} \textbf{a} Cells navigate a chemoattractant landscape $ S(x) $ using a run and tumble strategy with characteristic scales and variables as represented in the picture ($\ell_0$, $\lambda_0^{-1}$ and $v_0$ are the typical run length, run time and ballistic run speed respectively). The simulations are run in a long domain of length $\widetilde{L} \gg \ell_0$ over long times $\widetilde{T} \gg \lambda_0^{-1}$. 
\textbf{b} The swimming cell senses the attractant concentration along its trajectory $x(t)$ and modulates its tumbling rate $\lambda(S(x(t)))$ by the chemotaxis transduction pathway with response given by Eq. \eqref{tumblebias}. The dynamic response of the pathway is mediated by the response kernel~$K(t)$ (Eq.~\eqref{eq:kernel}).
\textbf{c} The shape of the bi-lobed kernel $K(t)$ normalised by its amplitude $K_\text{max}$ against time. 
}
\label{Fig1}
\end{figure}

A classical setup for chemotaxis is represented schematically in Fig.~\ref{Fig1}a. Cells swim following a run-and-tumble motion: ballistic  motion (`runs') at constant velocity $ v_0 $, interrupted by random re-orientations (`tumbles') occurring at random times governed by a Poisson process with rate $ \lambda(t) $~\cite{othmer}. As cells swim along their trajectory $ x(t) $ (taken here to be one-dimensional for simplicity), they are exposed to an attractant concentration $S(x(t))$. Assuming initial adaptation to the ambient attractant concentration, the cells modulate their Poisson tumbling rate according to~\cite{tu,deGennes}: 
\begin{equation} \label{tumblebias}
\begin{aligned}
 &\lambda (t) = 1-\Lambda(t) \\
 &\text{with} \quad \Lambda(t) = \int_{-\infty}^{t} K(t-u) S(x(u))d u \,.
 \end{aligned}
\end{equation} 
This represents the dynamics of the chemotaxis signal transduction, illustrated on Fig. \ref{Fig1}b.
Throughout, we use variables  non-dimensionalised with respect to the characteristic length and time scales: 
\begin{equation*}
x=\frac{\widetilde{x}}{\ell_0}, \enskip t=\lambda_0 \widetilde{t}, \enskip \lambda=\frac{\widetilde{\lambda}}{\lambda_0}, \enskip 
S=\frac{\widetilde{S}}{S_\text{tot}},
\end{equation*}
where  $ S_\text{tot} $ is the total attractant concentration, $\lambda_0=1\, s^{-1}$ is the basal tumbling rate, and $\ell_0 = v_0/ \lambda_0=10\, \mu m$ is the average run length~\cite{berg:72}. 

In Eq.~\eqref{tumblebias}, $ K(t) $ is the chemotactic memory kernel, measured through impulse response experiments~\cite{berg:86}, which has a bi-lobed shape for some attractants in~\textit{E. coli}~\cite{clark,celani} (Fig.~\ref{Fig1}c). A typical form for $K(t)$  is given by:
\begin{equation}
\label{eq:kernel}
K(t) = \frac{\beta}{\gamma} e^{- t/\gamma} \left(\frac{  t}{\gamma}-\frac{t^2}{2\gamma^2}\right),
\end{equation}  
where $ \beta $ is a dimensionless signal gain, and $\gamma =  \lambda_0 \widetilde{\gamma}$ is the cellular memory, a (dimensionless) relaxation time, as seen by the fact that the crossing point of the bi-lobed response is $t= 2 \gamma$ (Fig.~\ref{Fig1}c). Note that the amplitude of the response kernel is $K_\text{max} =  (\sqrt{2}-1) e^{\sqrt{2}-2}  \beta/\gamma$. Hence an increase in memory decreases the overall response. The kernel in Eq.~\eqref{eq:kernel} can be understood as a linear filter with three states (Supplementary Note 1), with a topology that achieves perfect adaptation~\cite{doyle,elsamad}, since $\int_0^\infty K(t) dt = 0$. Our setup differs from an alternative model~\cite{dufour} based on the dynamics of the CheY protein regulating the tumbling rate, which typically leads to a one-state linear filter obtained through linearisation. In Supplementary Note 2, we show that this alternative model can be equivalently written in the form of Eq.  \eqref{tumblebias} but with a kernel $K(t) = \beta(\delta(t) - (1/\gamma)e^{-t/\gamma})$ with a singularity at $t=0$ and a single decaying exponential, instead of the bi-lobed kernel (Eq.~\eqref{eq:kernel}).

At long time scales involving many runs and tumbles ($ t \gg 1 $), the swimming behaviour may be approximated by a drift-diffusion process~\cite{rousset,othmer2,xue}. 
In this regime, the time evolution of the population density of cells $ \rho(x,t) $ from an initial state $ \rho(x,0)$ is described by a Fokker-Planck partial differential equation: 
\begin{equation}\label{KS}
\frac{\partial \rho}{\partial t} - \frac{1}{2}\frac{\partial^2\rho}{\partial x^2} +  \frac{\partial}{\partial x} (\rho v) = 0,
\end{equation}
where $v(x,t)=\widetilde{v}(x,t)/v_0$ is the drift velocity of the cells, and the diffusion coefficient $D=\ell_0^2 \lambda_0$ drops out as part of the non-dimensionalisation. Equivalently, $\rho(x,t) $ is the probability of finding a cell at $x$ after time $t$ from a starting position $x_0$ drawn from $ \rho(x,0)$.

Typically, derivations in the literature \cite{rousset,erban,othmer} consider the regime of long memory (compared to the average run) and shallow perceived gradient (i.e., the attractant does not vary appreciably over the memory):
\begin{equation} 
\label{eq:sep_timescales}
1 \ll \gamma \ll \left(\beta\frac{\partial S}{\partial x} \right)^{-1}.
\end{equation}
Under these assumptions, the drift velocity $ v(x,t) $ can be shown~\cite{deGennes} to align with the local gradient (see Supplementary Note 3):
\begin{equation}\label{vKS}
v(x,t) =   \chi \frac{\partial S}{\partial x} =: v_\text{KS}(x), 
\end{equation} 
where the 
chemotactic response coefficient $\chi$ follows from the kinematics and the memory kernel (Eq.~\eqref{eq:kernel}):
\begin{equation}
	\chi =\frac{2\beta \gamma }{(1+2 \gamma )^3 }\,. \label{eq:Chi} 
\end{equation}
Eq.~\eqref{KS}, together with Eqs. \eqref{vKS} and \eqref{eq:Chi}, defines the classic linear Keller-Segel (KS) equation for the time evolution of the population density under a landscape $ S(x) $. We denote the solution to this equation as $ \rho_\text{KS}(x,t;\,S) $.

However, the KS model is actually valid under the weaker condition~\cite{xue}
\begin{equation}  
|\Lambda| \ll 1 \quad\textit{(small response)}, \label{smallresp}
\end{equation} 
i.e., the tumbling response remains close to the adapted value.
It can be shown that time scale separation (Eq.~\eqref{eq:sep_timescales}) implies small response (Eq.~\eqref{smallresp}), but the converse is not necessarily true. Hence KS can still be valid in the realistic situation when Eq.~\eqref{eq:sep_timescales} breaks down because the cellular memory is commensurate with environmental fluctuations~\cite{becker,aquino,andrews}, as long as Eq.~\eqref{smallresp} holds. Below, we consider a broad span of memory values (from the well separated to the commensurate) but always in the small response regime so that the KS model is valid.

\section*{Chemotaxis of cells with memory studied using agent-based numerics} 

We consider cells with memory swimming in a rugged environment with spatial correlations, leading to a temporally fluctuating input perceived along their trajectories. To study the effect of memory, we performed agent-based (AB) simulations of run-and-tumble motion as in Refs.~\cite{sontag,becker} coupled to a cellular response (Eqs.~\eqref{tumblebias}--\eqref{eq:kernel}) with memory (see Supplementary Note 1 for details). 

Our rugged landscape is a simple linear attractant concentration profile with additive spatial noise~\citep{becker}:
\begin{equation}
\label{Seta}
S_{\eta}(x) = \alpha \, x + \eta(x),
\end{equation}
where $\eta(x) $ is a random spatial variable described by the stochastic harmonic oscillator Langevin equation:
\begin{equation}
\label{eq:SHO}
\begin{aligned}
\frac{d}{dx}\eta(x) &= \theta(x)\\
m \frac{d}{dx}\theta(x) &= -\frac{1}{\mu} \eta(x) -  \theta(x) + \sigma_\eta \sqrt{\frac{2}{\mu}} \, \xi(x).
\end{aligned}
\end{equation}
Here $ \xi (x) $ is a unit white noise, and 
\begin{equation*}
\alpha = \frac{\ell_0}{S_\text{tot}} \, \widetilde{\alpha}, \enskip  \sigma_\eta=\frac{\widetilde{\sigma}_\eta}{S_\text{tot}} , \enskip 
m = \frac{ \widetilde{m}}{\ell_0}, \enskip \mu = \frac{\widetilde{\mu}}{\ell_0}
\end{equation*}
are non-dimensionalised parameters corresponding to: attractant gradient, noise variance, inertia, and spatial correlation length, respectively. 
This random landscape has two desirable properties. First, $ S_\eta(x) $  with $m>0$  is continuous and differentiable, so that
\begin{equation}
\label{eq:eta_differentiable}
 \left \langle \frac{\partial S_\eta}{\partial x} \right \rangle_\xi = \alpha + \left \langle  \frac{\partial \eta}{\partial x} \right \rangle_\xi = \alpha, 
\end{equation}
where $\langle\cdot\rangle_\xi$ denotes averaging over independent realisations of $\eta(x)$.
Second, Eq. \eqref{eq:SHO} is a regularised spatial Ornstein-Uhlenbeck (OU) process (Supplementary Figure 1): as $m \to 0$, $\eta$ converges to an OU process $\eta^0(x)$ which has exponential correlations with characteristic length $\mu$: 
\begin{equation}
\label{eq:correlations}
\langle \eta^0(x)\eta^0(x') \rangle_\xi = \sigma_\eta^2e^{-|x-x'|/\mu}=:C_\eta(|x-x'|).
\end{equation}
The OU limit is used below to facilitate our analytical calculations.

\section*{Chemotaxis in constant shallow gradients}

We first consider the landscape with zero ruggedness:
\begin{equation}
S_0 = \alpha \, x,  
\label{eq:gradient_nonoise}
\end{equation}
which corresponds to $\sigma_\eta =0$ or, alternatively, to the limit $\mu  \to \infty$, when the correlation length diverges. 
In this case, it has been shown~\cite{erban,othmer2} that the condition  
\begin{equation}\label{smallab}
\beta\alpha \ll 1\quad \textit{(shallow perceived gradient)}  
\end{equation}
guarantees that the small response condition (Eq.~\eqref{smallresp}) also holds. Hence we expect the AB numerics to be well described by the KS equation.

\begin{figure} 
	\centering
	\includegraphics[width=\columnwidth]{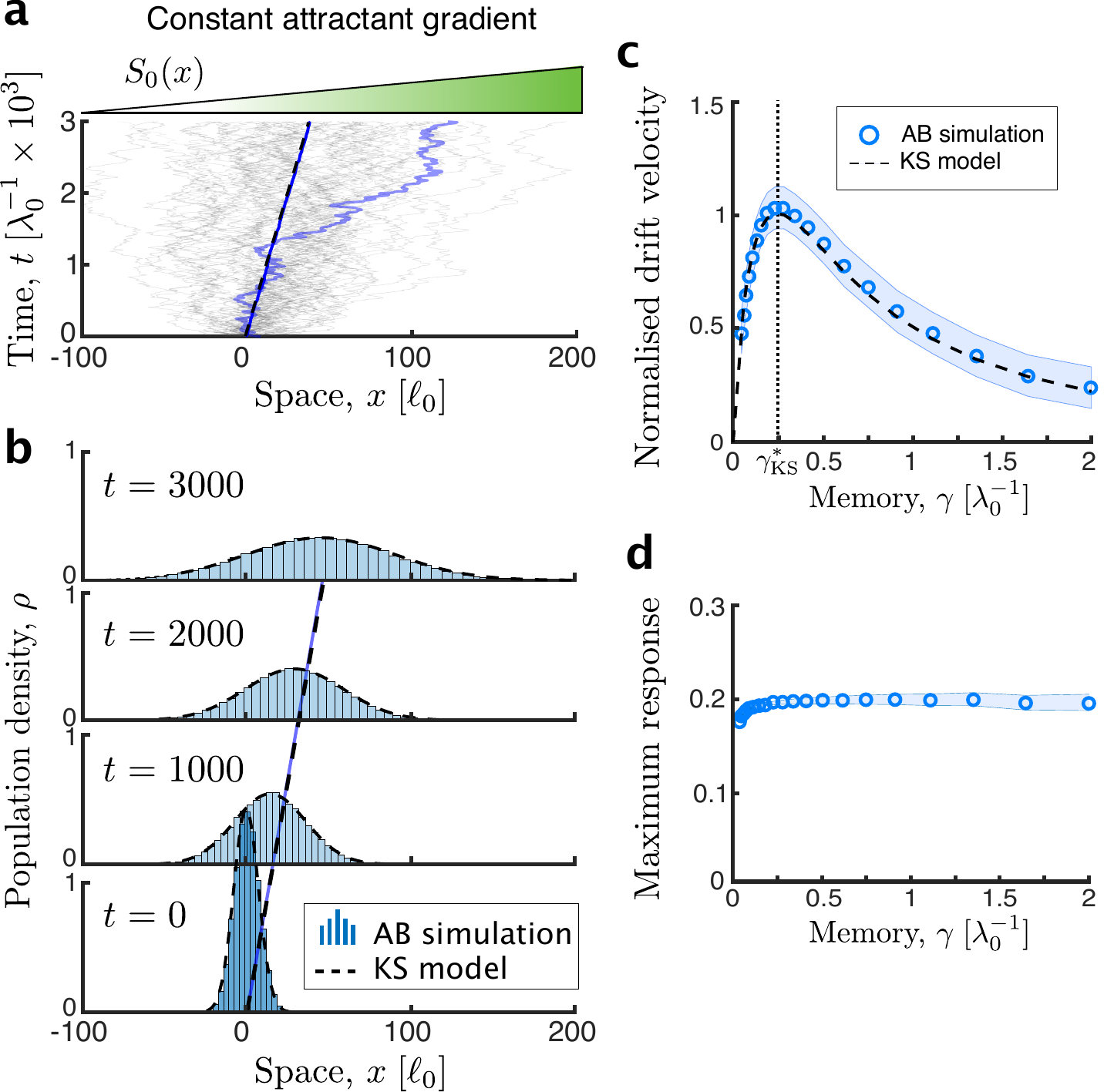}
	\caption{ \textbf{Comparison of agent-based numerics and Keller-Segel approximation in shallow gradients.}
	The AB model is used to produce $N =10^5$ cell trajectories over $T = 4\times 10^3$ ($\Delta x = 5\times 10^{-5}$, $\Delta t =5\times 10^{-3}$) with perceived gradient $\beta \alpha=0.1$, $\beta \sigma_\eta = 0$. The KS model is integrated numerically using a first-order in time, second-order in space forward-Euler scheme ($\Delta x = 10^{-4}$, $\Delta t = 1$).
	\textbf{a} Sample trajectories of the AB model ($\gamma = 0.5$) in the deterministic landscape $ S_0(x) $. 
	\textbf{b} The evolution of the population density of the AB model ($\rho_\text{AB}(x,t;\, S_0)$, histogram) is well captured by the evolution of the KS equation ($ \rho_\text{KS}(x,t;\,S_0) $, Eq.~\eqref{KS}, dashed line, $\gamma = 0.5$). The solid blue line indicates the average velocity $\vAB$, which is indistinguishable from the KS drift $\vKS$. The densities are normalised to unit mass.
	\textbf{c} The drift velocity from the AB model ($ \vAB $, circles) is well predicted by the KS drift velocity ($ \vKS $, dashed line). Both velocities are normalised by the maximal KS drift velocity $v_\text{KS}^*$, which is reached at a memory of $\gamma_\text{KS}^*=1/4$, shown by the dotted line.
	\textbf{d} The maximum tumbling response $\max|\Lambda|$ (circles) stays well below unity, showing that the small response condition is met for all the simulations. The blue band indicates the standard deviation of the simulations.
	} 
	\label{Fig3}
\end{figure}

To test this prediction, we used the AB model to simulate $N=10^5$ independently generated cell trajectories $\{x_\text{AB}(t; \,S_0);\; t\in(0,T) \}$, where $T=4 \times 10^3$ (Fig. \ref{Fig3}a), from which we obtain population snapshots, $\rho_\text{AB}(x,t; \,S_0)$. All the simulations were run in the regime of small $\beta \alpha$.
In Fig.~\ref{Fig3}b, we show that the statistics of the AB simulations are well captured by the continuum KS solution:
\begin{align*}
\rho_\text{AB}(x,t; \,S_0) \approx \rho_\text{KS}(x,t; \, S_0).
\end{align*}
We also compared the drift velocity of the KS solution to the average velocity of AB cells (computed over the long simulation time $T$): 
\begin{equation}\label{vdrift}
\begin{aligned}
	&\vAB(S_0) :=\left\langle   \frac{  x_\text{AB}(T;S_0) -   x_\text{AB}(0; S_0) }{T} \right\rangle_\text{AB} \\ 
	&\vKS (x; S_0) = \chi \frac{\partial S_0}{\partial x} = \chi \,\alpha,
\end{aligned}
\end{equation}
where $ \langle\cdot\rangle_\text{AB} $ denotes averaging over the ensemble of AB cells. Fig.~\ref{Fig3}c shows that the average velocity of the AB population matches the drift velocity of the KS model for varying memory $\gamma$. 

Maximising Eq. \eqref{eq:Chi} shows that the drift velocity $\vKS$ achieves a maximum at an optimal memory:
\begin{equation}
\vKSopt := \max_\gamma \vKS(S_0)  \quad \text{at $ \gamma_\text{KS}^\ast= 1/4 $}, 
\label{KSmax}
\end{equation}
Hence a cell with optimal memory $\gamma_\text{KS}^\ast$ has a kernel $K(t)$ with a zero crossing at $t=1/2$, i.e., halfway through the expected length of a run (see Fig. \ref{Fig1}c). KS thus predicts that the drift speed is maximal when the gradient is measured along a single run, when the cell can take an unbiased measurement while moving in a straight line. 
For the zero-ruggedness landscape, our AB simulations (Fig.~\ref{Fig3}c) also display a maximum in the average velocity of the population when the cells have memory $\gamma = \gamma_\text{KS}^\ast$.

Fig.~\ref{Fig3}d confirms that the simulations are in the regime of small response (Eq.~\eqref{smallresp}) where KS holds.
As $\beta \alpha$ is increased, and the small response condition (Eq. \eqref{smallresp}) is violated, the correspondence between the AB and KS solutions gradually breaks (see Supplementary Figure 2).

\section*{Chemotaxis in rugged, correlated landscapes}

The kernel $K(t)$ with intrinsic memory $\gamma$ has been shown to filter high-frequency input noise~\cite{aquino,becker}. 
However, cells could also use this memory to their advantage as they process the correlated fluctuations that they encounter as they traverse a rugged landscape.

To test this idea, we carried out AB simulations of cells with memory navigating the spatially correlated landscape (Eq.~\eqref{Seta}) and compared it to the predicted KS behaviour. To ensure that the differences between AB and KS are a direct consequence of the correlated spatial fluctuations, all our simulations are run in the small response regime (Eq.~\eqref{smallresp}) where KS holds, while keeping a large signal-to-noise ratio (Supplementary Note 4): 
\begin{align}
\label{signaltonoise}
\alpha \gg \sigma_\eta \quad\textit{(large signal-to-noise ratio)},
\end{align}

\begin{figure}[ht!]
	\centering
	\includegraphics[width=\columnwidth]{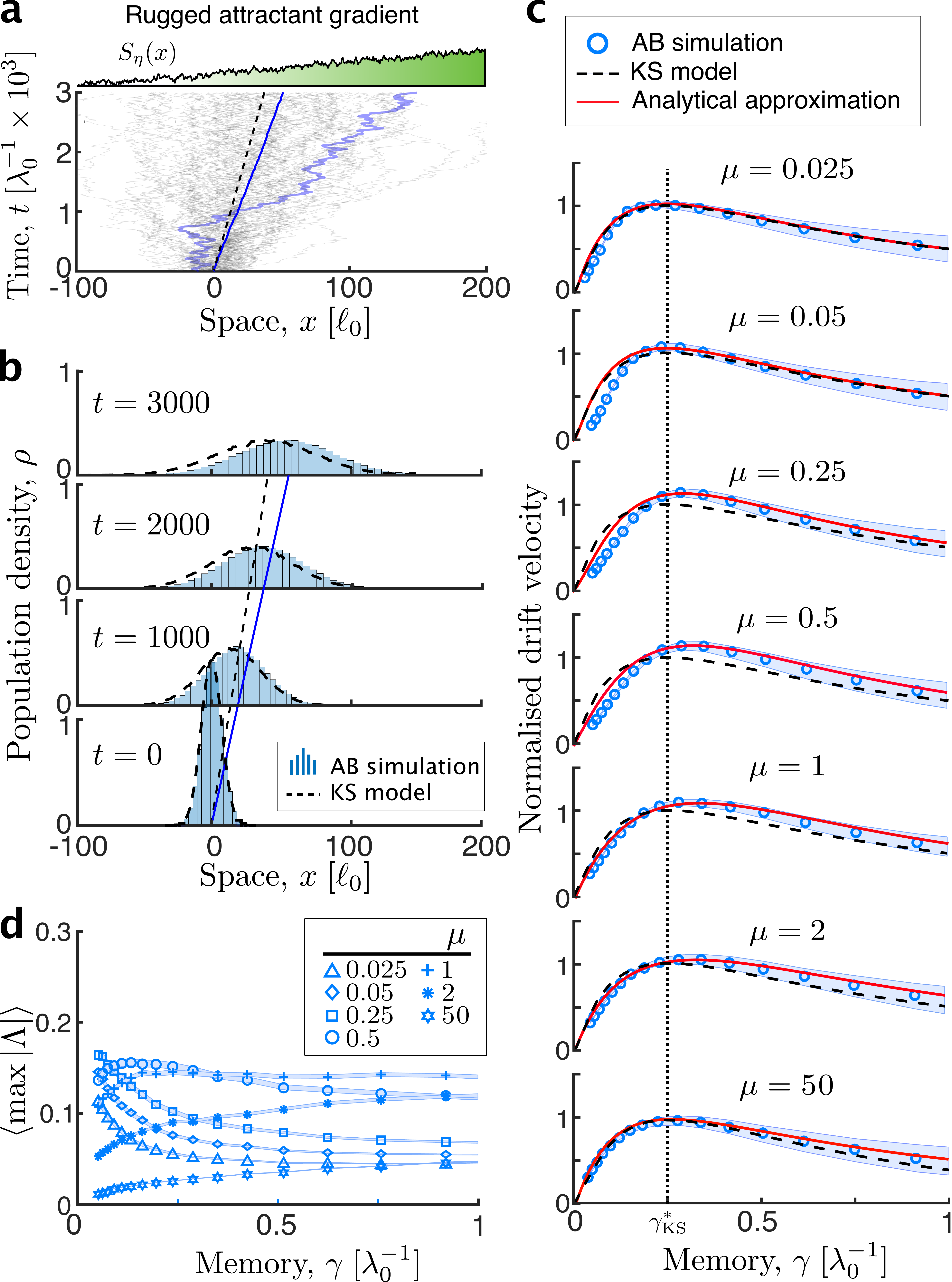}
	\caption{ \textbf{Comparison of agent-based simulations and analytic predictions in rugged landscapes $S_\eta$.}
	The AB model is used to produce $N =10^4$ cell trajectories over $T = 4\times 10^3$ ($\Delta x = 5\times 10^{-5}$, $\Delta t =5\times 10^{-3}$) in $ 10^2 $ realisations of $ S_\eta(x) $ with perceived gradient $\beta \alpha=0.05$, $\beta \sigma_\eta = 10^{-3}$. The KS model is integrated numerically using a first-order in time, second-order in space forward-Euler scheme ($\Delta x = 10^{-4}$, $\Delta t = 1$).
	\textbf{a} Sample trajectories of the AB model ($\gamma = 0.5$) in the rugged landscape ($ \mu = 1 $). Note that the ensemble average drift of the AB cells (solid blue line) is faster than the KS drift (dashed black line).  
	\textbf{b} The evolution of the population density of the KS model ($\langle\rho_\text{KS}(x,t;\, S_\eta) \rangle_\xi$, Eq. \eqref{KS}, dashed black line, $\gamma = 0.5$, $ m = 5 \times 10^{-3} $) generally fails to capture the evolution of the AB population ($\langle\rho_\text{AB}(x,t;\, S_\eta) \rangle_\xi$, histogram). The densities are normalised to unit mass.
	\textbf{c} The average drift velocity of the AB model ($ \vABnoise$, circles) as a function of the memory $\gamma$ is well described by our approximation ($\Vmu $, Eq.~\eqref{Vmu}, red solid line) for all values of $\mu$, but in general not well captured by the KS drift ($ \vKS $, Eq. \eqref{vKS}, black dashed line). The AB drift is averaged over realisations of the rugged landscape and both velocities are normalised by the maximal KS drift $v_\text{KS}^*$ in the deterministic landscape.
	\textbf{d} The average maximum response amplitude $|\Lambda|$ from the AB simulations at different values of $\mu$ (different symbols in the figure) remains much smaller than unity for all combinations of $\gamma$ and $\mu$ (as in Fig. \ref{Fig3}d), indicating that the KS model still holds in the rugged landscape. Yet, as panels b, c show, it does not capture the drift velocity. The blue bands indicate the standard deviation of the simulations. }
	\label{Fig4}
\end{figure}

Fig.~\ref{Fig4} shows that the AB cell population travels faster than predicted by KS going up the gradient of the rugged landscape.
Fig.~\ref{Fig4}a presents simulated AB trajectories for a particular realisation of the landscape $S_\eta(x)$, and Fig.~\ref{Fig4}b compares the time evolution of the KS solution $\langle \rho_\text{KS}(x,t;\, S_\eta) \rangle_\xi$ to the empirical distribution from the AB numerics $\langle \rho_\text{AB}(x,t;\, S_\eta) \rangle_\xi$, both averaged over $10^2$ independent realisations of the landscape $S_\eta(x)$. Our numerics show that the AB distribution propagates faster: $\vABnoise > \vKS$, i.e., the average cell velocity of the AB simulations (defined in Eq.~\eqref{vdrift}) averaged over realisations of the landscape (blue solid line) is larger than the corresponding KS drift velocity (dashed line).

We have examined this enhanced chemotaxis as a function of the length scale of the landscape.  We show in Fig.~\ref{Fig4}c that for correlation lengths around the run length ($\mu \approx 1$), the KS drift velocity $ \vKS$ underestimates the average velocity $\vABnoise$ of cells with memory $\gamma \geq \gamma_\text{KS}^\ast= 1/4 $.

As expected, the average velocity of the AB cells is well approximated by the KS solution in the limits of both vanishingly small and infinitely large correlation length:
\begin{equation}
  ||\vABnoise -  \vKS|| \to 0 \quad \text{as} \quad  \mu \to \{0,\,\infty\},  \label{ABKSlim}
\end{equation}
which correspond to an uncorrelated landscape or a zero-ruggedness constant gradient, respectively. Also expected, Eq. \eqref{ABKSlim} holds in the memoryless limit $\gamma\to 0$. In this limit, the kernel $ K(t) $ computes the temporal derivative of the signal (see Supplementary Note 1), and the tumble rate (Eq.~\eqref{tumblebias}) is  $\lambda(t) \simeq 1-\beta d S/dt$, so that the drift velocity is given by Eq.~\eqref{vKS} (see Sect. 2.3 in Ref. \cite{rousset}). This result is consistent with fast adaptation dynamics approaching gradient-sensing~\cite{sontag,rousset}.

Note that the system is in the small response regime (Eq. \eqref{smallresp}) where KS is applicable (compare Fig.~\ref{Fig4}d with Fig.~\ref{Fig3}d). Yet, our AB numerics show that cells with memory can drift faster than predicted by mere gradient sensing (KS) when navigating  environments with spatially correlated fluctuations, thus indicating a role for cellular memory in using spatial information beyond local gradients.

\section*{Derivation of drift speed in rugged landscapes}

To capture the numerically observed enhancement of AB chemotaxis in rugged landscapes, we extend de Gennes' analytical derivation of the drift velocity to incorporate the interaction of memory with the landscape fluctuations. To facilitate our analysis, in the rest of this section we work in the OU limit of the landscape, i.e., $m \to 0$ in Eq.~\eqref{eq:SHO}.

Consider a population of cells navigating the rugged landscape $ S_\eta(x) $. Under chemotaxis, the average duration of runs up the gradient $\left\langle t^+ \right \rangle_\text{AB}$ is larger than the duration of runs down the gradient $\left\langle t^- \right \rangle_\text{AB}$.  Following de Gennes~\cite{deGennes}, it can be shown that the (non-dimensionalised) average velocity of the cells is:
\begin{align}
	v_\text{AB}(S_\eta) &= 	\frac{\left\langle t^+ \right \rangle_\text{AB} - \left\langle t^- \right \rangle_\text{AB}}{\left\langle t^+ \right \rangle_\text{AB} + \left\langle t^- \right \rangle_\text{AB}} = \left\langle t^+ \right \rangle_\text{AB} - 1,
	\label{vball}
\end{align}
where we use the fact that $\left\langle t^+ \right \rangle_\text{AB} + \left\langle t^-  \right \rangle_\text{AB} = 2$, i.e., the sum of the average duration of one up-hill and one down-hill run is equal to two runs. Equation~\eqref{vball} just states that the average cell velocity is the excess duration of the average up-gradient run beyond the duration of an average run. 

Using the ergodicity of the run-and-tumble process, the expectation over AB trajectories becomes a time integral: 
\begin{align}\label{trun}
	\langle t^+\rangle_\text{AB} &= \int_0^\infty s \, p(s\,|\,x(t))\,ds = \int_0^\infty  e^{ -\int_0^s\lambda(t) dt} \,ds, 
\end{align}
where $p(s \,|\, x(t)) =  \lambda(s)\exp ( -\int_0^s\lambda(t) dt)$ is the conditional probability density of Poisson tumble times given the path $x(t)$. In the small response regime (Eq.~\eqref{smallresp}), we can expand the exponential to second order to obtain
{\small
\begin{align*}
\langle t^+\rangle_\text{AB} & \simeq 
1 + \int_0^\infty e^{-s} 
\left [\int_0^s \Lambda(t) dt + \frac{1}{2} \left(\int_0^s \Lambda(t) dt \right)^2          \right ] ds .
\end{align*}
}

We now depart from de Gennes' derivation and consider
the cell velocity (Eq.~\eqref{vball}) averaged over realisations of the landscape $S_\eta$:
\begin{align}
\vABnoise \simeq & 
\int_0^\infty e^{-s} \left\langle \int_0^s \Lambda(t) dt \right\rangle_\xi ds \nonumber \\
+& \frac{1}{2} \int_0^\infty e^{-s} 
 \left\langle \left(\int_0^s \Lambda(t) dt \right)^2  \right\rangle_\xi  ds ,
\end{align}
where the first term does not depend on the spatial noise:
\begin{align}\label{firstterm}
\left\langle\int_0^s\Lambda(t) dt \right\rangle_\xi &= 
\int_0^s \int_0^\infty K(u) \left\langle S_\eta(x(t-u)) \right\rangle_\xi du dt 
\nonumber \\
&= \int_0^s \int_0^\infty K(u) S_0(x(t-u)) du dt,
\end{align}
and the second term contains the effect of the spatial correlations as a result of the overlap between the memory kernel and the spatial covariance (Eq.~\eqref{eq:correlations}): 
{\small
\begin{align}
&\left\langle\left(\int_0^s\Lambda(t)dt\right)^2\right\rangle_\xi= \left(\int_0^s\int_0^\infty  K(u)S_0(x(t-u)) du dt \right)^2 \nonumber\\
&+  \int_0^\infty K(w)\int_w^s \int_{w-\widehat{t}}^0\left( \int_0^\infty K(u) \, C_\eta(x(\tau-u))\,du \right)d \tau d \widehat{t}\,dw .
\label{term22}
\end{align}
}
Here $\tau = t-\widehat{t}+w$ represents the delay between the input $\eta^0(x(t-\tau))$ and the output $\Lambda(t)$, and the limits of integration reflect causality. 

Collating Eqs.~\eqref{trun}--\eqref{term22} and integrating, we obtain our approximation of the drift velocity in rugged landscapes: 
\begin{equation}\label{Vmu}
\vABnoise \simeq
\vKS +\vcorr =: \Vmu , \\ 
\end{equation}
where Eq.~\eqref{firstterm} gives rise~\cite{deGennes} to the KS drift velocity (Eq.~\eqref{vKS}):
\begin{align}
    \vKS=\beta \alpha \, \frac{2 \gamma }{(1+2 \gamma )^3 },
    \label{eq;vKS_explicit}
\end{align}
and Eq.~\eqref{term22} leads to the correction due to spatial correlations:  
{\small
\begin{align}
\label{eq:correction}
\vcorr 
&= \frac{\beta^2 \sigma_\eta^2}{2}\frac{ \gamma ^2 \mu  \left[2 \gamma ^3 (1+\mu) + (1+\gamma)^3 \mu ^2  + 6 \gamma ^2 \mu  -2  \mu ^2\right]  }{(1+\gamma )^6 (1+\mu ) (\gamma +\mu )^3}
\end{align}
}
For further details, see Supplementary Note 5.  

Our approximation $\Vmu$ recovers the KS drift velocity in different limits: for deterministic and uncorrelated landscapes ($\Delta v_\mu \to 0 \text{ as } \mu\to \{0,\,\infty\}$); in the zero and infinite memory limits ($\Delta v_\mu \to 0 \text{ as }\gamma\to \{0,\,\infty\}$); as well as in the limit of vanishing gradient ($\Vmu \to \vKS \text{ as } \alpha\to 0$), since (Eq.~\eqref{signaltonoise}) is required to derive $\vcorr$.

\begin{figure}
	\centering
	\includegraphics[width=0.8\columnwidth]{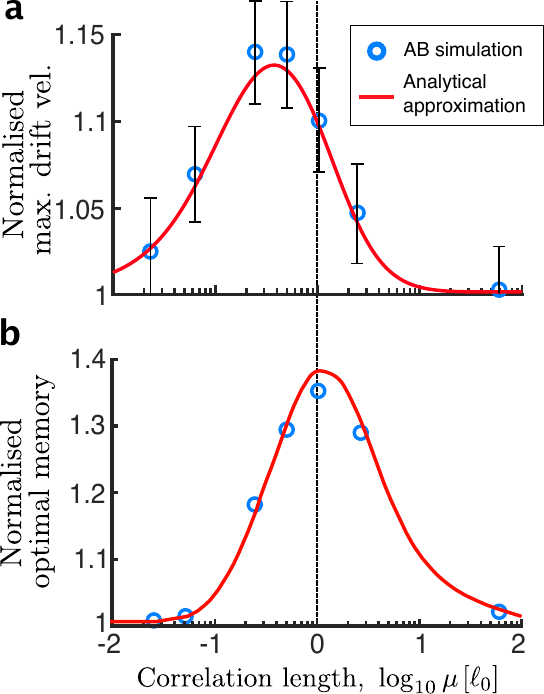}
	\caption{ \textbf{Dependence of the maximal drift velocity and the optimal memory on the environmental correlation length.}
	\textbf{a} The maximum drift velocity from the agent-based (AB) numerics at various correlation lengths of the landscape (circles) is well predicted by our approximation (solid line). This maximum, achieved in rugged gradients, exceeds the corresponding deterministic Keller-Segel (KS) drift for all $\mu$, and approaches the KS prediction in the white noise ($\mu\to 0$) and constant gradient ($\mu\to \infty$) limits. The drift velocities are normalised with respect to the maximum attainable KS drift in deterministic landscapes. The error bars indicate the standard deviation of drift velocity fluctuations across realisations of the rugged landscape at the optimal memory $\gamma^*_\mu$ corresponding to this drift speed.
	\textbf{b} The optimal memory against the correlation length shows that the memory at which the drift velocity is maximised is always larger in AB cells than the optimal KS value: $ \gamma^*_\mu/\gamma_\text{KS}^* \ge 1 $. The longest optimal memory for a given correlation length occurs for $\mu = 1$ (dotted line), which equals the expected run length. Thus, the tumble rate limits the length scale of perceivable fluctuations.
	} 
	\label{Fig5}
\end{figure}

\section*{The effect of memory on the drift speed }

Our approximation (Eq.~\eqref{Vmu}) makes explicit the fact that cells use both the local gradient (through $\vKS$) and the spatial correlations (through $\vcorr$) to navigate rugged landscapes. Fig.~\ref{Fig4}c shows that, in contrast to the KS drift, our $ \Vmu $ predicts the enhanced cell velocity in the AB simulations, $\vABnoise$, and its dependence on the landscape length scale $\mu$ for a broad range of memory,~$\gamma$.

Fig.~\ref{Fig5} compares the predicted maximal velocity and the optimal memory at which it is achieved,
\begin{equation}\label{maxV}
\Vmuopt = \max_\gamma \Vmu  \quad \text{attained at $ \gamma_\mu^\ast $}
\end{equation}
with the numerical simulations. As expected, we recover the KS behaviour in both limits of deterministic ($\mu \to \infty$) and uncorrelated ($\mu \to 0$) landscapes, when there is no advantage in using memory to use the statistical correlations of the environment. The optimal memory $ \gamma_\mu^\ast $ thus emerges as a balance between filtering and tumbling: for $\gamma > \gamma_\mu^\ast$, cells improve their noise filtering~\cite{becker} but lose orientation due to the larger number of tumbles taken account in their history; on the other hand, for $\gamma < \gamma_\mu^\ast$, cells are less likely to tumble, but filtering of environmental noise becomes suboptimal. Our results in Fig.~\ref{Fig5}a show that the velocity of cells with optimal memory is always larger than the gradient-sensing KS drift velocity (i.e., $\Vmuopt(\mu) \ge \vKSopt, \enskip \forall \mu$), with the largest improvement at $\mu \simeq 1/2$, the point where the length scale of the environmental correlations are on the order of one half of a run length. As seen in Fig.~\ref{Fig5}b, the corresponding optimal memory is also always larger than the KS memory: $\gamma_\mu^\ast \ge \gamma^\ast_\text{KS}=1/4$. Our calculations show that it is advantageous to increase the memory when the correlations are of the same order as the length of the run ($\mu = 1$); yet for correlations longer than one run ($\mu \geq 1$), the presence of random tumblings erode this advantage and the optimal memory returns to the KS value..

This behaviour is consistent with models of tethered cells receiving noisy temporal stimuli ~\cite{becker}. In particular, the mutual information between input and output with a delay $\tau$ is maximised when maximising the correlation
{\small
\begin{align}
R_{\eta\Lambda}(\tau)&:= \frac{\langle\eta^0(t-\tau)\Lambda(t)\rangle_\xi}{\sigma_\eta\sigma_\Lambda}
=\frac{1}{\sigma_\eta\sigma_\Lambda}\int_0^\infty K(u) R_\eta(\tau-u)\,du\, ,
\label{eq:tenWolde_corr}
\end{align}
}
which is a normalised version of the integral in brackets in the second term of Eq.~\eqref{term22}. It was shown~\cite{becker,tostevin} that Eq.~\eqref{eq:tenWolde_corr} is maximised for a memory corresponding to optimal filtering, and, consistently, our results reflect the importance of noise filtering. For navigation, however, the optimality of filtering is not the only criterion, and it needs to be balanced with the random tumbling time scale which imposes a threshold on the bandwidth of correlations that are useful to improve drift speed.

\section*{The effect of memory on the heterogeneity of the population}

Thus far, we have shown that our approximation predicts well the effect of memory on enhancing the cell velocity in rugged landscapes, although, as seen in Fig.~\ref{Fig4}c, it overpredicts the velocity of AB cells with short memory ($\gamma < \gamma^*_\text{KS}=1/4$) navigating mildly rugged landscapes ($0.05 < \mu < 0.5$). The origin of this discrepancy is in the fact that cellular memory has an effect not only on the average cell velocity but also on the heterogeneity of the distribution of AB cells, an effect that is not captured by our approximation in Eq.~\eqref{Vmu}.

To characterise this behaviour further, we carry out additional numerical computations. Intuitively, we expect that cells with short memories will be more sensitive to local irregularities, and hence more prone to becoming disoriented in rugged landscapes. At the population level, this could lead to the appearance of subpopulations of propagating agents. On the other hand, cells with long memory will average their responses over extended patches of the landscape, thus being less sensitive to local fluctuations of the landscape and maintaining the unimodality of the distribution. 

\begin{figure}[t]
\centering
\includegraphics[width=0.95\columnwidth]{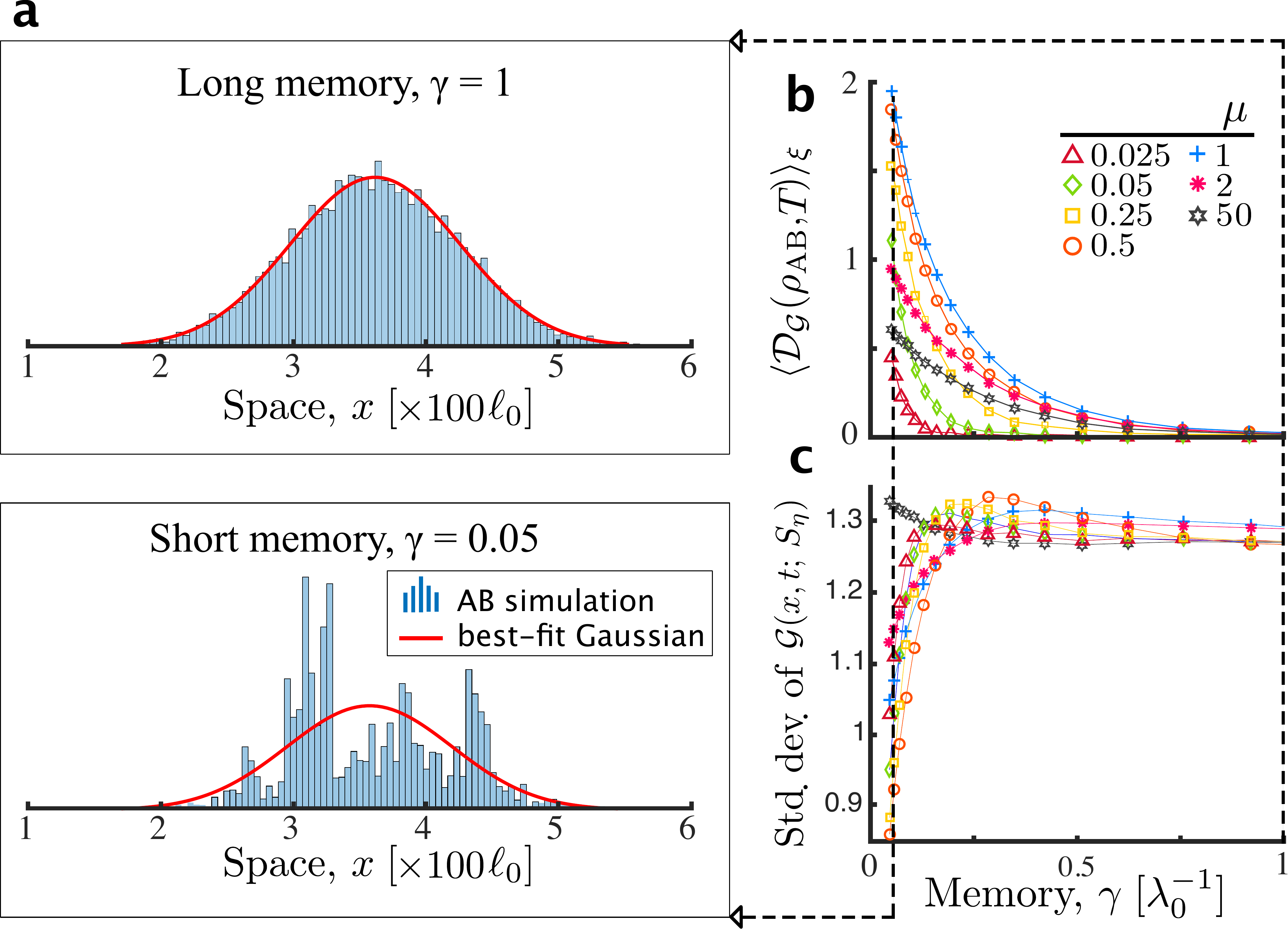}
\caption{\textbf{The cellular memory controls the population heterogeneity. }
\textbf{a} Snapshots of the agent-based population density $ \rho_{AB}(x,T;\,S_\eta) $ in a rugged landscape $S_\eta$  ($\beta\sigma_\eta=10^{-3}$, $\mu = 1$) measured at $T= 4\times 10^3 $ for two values of the memory $\gamma$ (histogram) shown with the best-fit Gaussian curves $ \mathcal{G}(x,T;\,S_\eta) $ (red line). For short memories (where short depends on the correlation length $\mu$) the population loses coherence.
\textbf{b} As $\gamma$ increases, the population becomes increasingly unimodal Gaussian. This is shown by $\mathcal{D}_\mathcal{G}(\rho_\text{AB},\,T)$, the distance between the population distribution and the best-fit Gaussian (Eq.~\eqref{eq:L2distance}) averaged over realisations of the landscape, decreasing to zero.
\textbf{c} For large $\gamma$, the standard deviation of $ \mathcal{G}(x,T;\,S_\eta) $ becomes independent of $\mu$ and $\gamma$, indicating that  randomness arises only from tumbling. 
}
\label{Fig6}
\end{figure}

A numerical illustration of this behaviour is presented in Fig.~\ref{Fig6}a, where we show the long-term AB numerics of two population of cells (one with long memories, another with short memories) starting from an initial Gaussian distribution and navigating a rugged landscape $S_\eta$. In the KS model, it is known that a Gaussian population remains Gaussian for all times~\cite{carrillo2}. Indeed, our AB numerics show that when cells have relatively long memories ($\gamma = 1$), the population does remain unimodal.
However,  the population of cells with short memories ($\gamma=0.05$) goes from being Gaussian to multimodal (i.e., with separate subpopulations), as time elapses. This behaviour is persistent over long times.

To quantify the loss of unimodality, in Fig.~\ref{Fig6}b we compute the $L_2$ distance 
between the AB distribution $\rho_{AB}(x,t;\,S_\eta)$ and its best Gaussian fit $\mathcal{G}(x,t; S_\eta)$ after a long simulation of $T=4 \times 10^3$: 
\small{
\begin{equation}
\label{eq:L2distance}
\left \langle \mathcal{D}_\mathcal{G}(\rho_\text{AB},T) \right \rangle_\xi =
\left \langle \left || \rho_\text{AB}(x,T;\,S_\eta)- \mathcal{G}(x,T; \,S_\eta) \right ||_{L^2}  \right \rangle_\xi \,.
\end{equation}
}
As discussed above, for a rugged landscape with length scale $\mu$, the AB distribution becomes increasingly Gaussian for larger $\gamma$,
converging to a Gaussian distribution (Fig.~\ref{Fig6}b-c). For large $\gamma$,  the standard deviation becomes largely independent of the landscape length scale $\mu$ (Fig.~\ref{Fig6}c). This is consistent with the fact that randomness arises not from the landscape but from tumbling, which is common to all cells, thus yielding unimodal population distribution in this limit. When the memory is short, on the other hand, cells use local information and their trajectories depend strongly on the starting positions, leading to distributions far from Gaussian. This dependence on the starting positions is reduced for cells with longer memory, which perceive and average overlapping information of the landscape.

These numerical results suggest that cellular memory could play a role not only in optimising long-term drift velocity (Fig.~\ref{Fig5}a) but also in controlling population level heterogeneity~\cite{frankel}. A trade-off between both objectives could then allow a more extended exploration of heterogeneous attractant landscapes by the cell population.

\section*{Discussion}

Chemotactic navigation relies on the processing of sensory information and its efficient transduction by the cellular mechanisms that control locomotion. Here we studied the run-and-tumble motion of \textit{E. coli} and investigated the role of the cellular memory inherently to the chemotactic response in determining the ability of cells to navigate rugged chemoattractant landscapes. In contrast to previous work, which focused on the deleterious effect of noise on sensing and swimming~\cite{endres,aquino}, our model explores how the correlations in the attractant, as encountered through swimming, can be used to enhance chemotactic performance. To this end, we considered an agent-based (AB) model of swimming cells with a history-dependent tumble rate computed by each cell along its trajectory through a signal-transduction model, and compared it to the classical Keller-Segel (KS) model, in which cells instantaneously align their velocity to the chemoattractant gradient independently of their history. Our results confirm that the KS model accurately predicts the behaviour of the AB population when navigating through constant, shallow chemoattractant gradients \cite{sontag,erban}. However, when the chemoattractant landscape has spatially correlated fluctuations superposed to the gradient, AB cells with short-term memory can exhibit greater drift velocity than predicted by mere gradient-alignment.  

Building on our numerical observations from the AB model, we extended work by de Gennes~\cite{deGennes} to derive an analytical approximation of the drift velocity that captures the ability of cells to use spatial correlations. The validity of our derivation hinges upon the assumption that the tumbling rate remains close to the adapted state~\cite{xue}. We show that this assumption, which is less restrictive than the standard shallow gradient assumption~\cite{erban,othmer2}, is enough to constrain the relationship between the time scales  of internal response and perceived stimulus. Specifically, our results hold for rugged environments in the large signal-to-noise regime, even if the typical shallow gradient assumption breaks down. Since our model only allows small tumble biases, the mechanism described here can be fully attributed to the linear filtering of perceived stimuli, and is therefore distinct from the non-linear response considered in previous work~\cite{dufour,long}. Because our analytic derivation of the drift velocity (Eq. \eqref{eq:correction}) links population dynamics to cell level parameters, it can be used to guide the experimental design to define parameter regimes where deviations from the Keller-Segel limit are expected.

Our analytical model predicts an enhancement of the drift velocity in rugged environments across a range of correlation lengths and cellular memories (Figs.~\ref{Fig4}--\ref{Fig5}). This result is consistent with cells with memory performing a non-local optimisation~\cite{ONpaper} beyond local gradient alignment. Importantly, when the landscape fluctuations are negligibly small or they occur on long spatial scales, our model recovers the KS model, so that the best strategy is purely local optimisation, as shown by previous studies~\cite{vergassola,celani}. Our findings are consistent with our fundamental understanding of bacterial chemotaxis: we find that cells relying only on instantaneous information tumble more often, leading to decreased average run length and drift velocity. Hence there is an ecological benefit for the cell to adjust their memory actively to match the length scales in the environment~\cite{sourjik}. We also show that our findings are in agreement with optimal information coding by the chemotactic pathway~\cite{endres,becker} and provide a link between previous results on memory and filtering in the time-domain~\cite{becker,andrews} to chemotactic spatial navigation.

Our model overpredicts the drift velocity of AB cells with short memory navigating mildly rugged landscapes (Fig.~\ref{Fig4}c). We showed that, in this regime, suboptimal filtering due to short memory results in the segregation of the bacterial population into long-lived multi-modal distributions (Fig.~\ref{Fig6}). This numerical observation suggests another distinct role for cellular memory as a means to tune how much the ruggedness of the landscape is reflected in the heterogeneity of the population responses. 

Micro-scale ruggedness in the landscape can have diverse origins. Ruggedness can appear naturally as a result of random spatial inhomogeneities associated with porous or particulate media, or due to degrading matter and secretions of other organisms. Such processes generate gradient structures on scales smaller or comparable to the mean run-length of bacteria~\cite{Blackburn,stocker}. In these environments, bacteria experience the spreading pulse of chemoattractant as a ruggedness in dilute, shallow background gradients~\cite{Hein,Blackburn}. There is evidence for this effect in native \textit{E. coli} environments: in soil, the length scales are estimated to vary between $\sim 2\,\mu \text{m}-1\text{ mm} $~\cite{Ngom}; in the digestive tract, the chemical environment is also thought to have fine-grained spatial structure~\cite{Hol}. Furthermore, some aquatic microbes also exhibit run-tumble motion, albeit at higher run speeds of $~\sim 100\,\mu\text{m }\text{s}^{-1}$ and with longer mean run lengths of $68–346\,\mu \text{m}$~\cite{watteaux}. Studies have found that the habitats of such microbes have inhomogeneities with length scales of $\sim 10-1000\,\mu\text{m}$, commensurate with the run lengths of bacteria. In addition, time-varying spatial ruggedness can also appear as a result of advection (stirring) of the medium, which causes it to form a network of thin, elongated filaments~\cite{stocker,torney} on scales of around $200–1000\,\mu \text{m}$, again commensurate with the run length of aquatic microbes.

Our one-dimensional model allows us to gain analytical insights into the relationship between memory and environmental fluctuations. In particular, we find that the optimal memory is always smaller than the run length. Hence cells are `blind' to longer correlations in one dimension, a fact that is consistent with short-term memory being typical of pre-adapted cells~\cite{berg:86}. 
In general, however, bacteria can optimise their adaptive responses (and hence their memory) from seconds to minutes through receptor methylation~\cite{sourjik}, suggesting the relevance of longer-scale memory-based sensing. This fact may be linked to the stronger directional persistence of bacterial motion in two or three dimensions.
With increased directional persistence, we expect that cells can further improve their drift speed by increasing their memory to exploit positive feedback from the environment during up-gradient runs, i.e., by eliciting large tumble rate reductions in comparison to small changes during down-gradient runs~\cite{tu,dufour,long}. However, understanding the interplay between memory and directional persistence would be most relevant when studied in conjunction with an appropriate description of 2D or 3D chemoattractant landscapes 
For instance, patchy landscapes, which are frequently considered in the ecological theory~\cite{Chang}, can give rise to complex navigation strategies~\cite{Hol} aimed at optimising the use of memory, swimming velocity and directional persistence. Future theoretical work should therefore move from quantifying chemotactic performance in terms of the drift speed in 1D landscapes towards studying swimming in 2D and 3D heterogeneous landscapes using a generalised performance measure, e.g., the amount of encountered attractant~\cite{ONpaper}.

Finally, although we have concentrated here on navigation in bacterial chemotaxis, memory-based search strategies are also relevant in other biological contexts for higher animals~\cite{ONpaper}. At a more conceptual level, other search and exploration optimisation processes could benefit from the consideration of agents with memory as a means to take advantage of spatial correlations. For instance, processes of visual search rely on specific neurons in the early visual cortex with a response function commonly approximated by a bi-lobed function (c.f. Fig. 6 in Ref. \cite{Adelson}). From this perspective, saccadic eye movements could be thought of as a navigation over the image with bi-lobed neurons using their memory to integrate the spatial correlations in the visual field as explored during the visual search.
Our work could also serve as inspiration for the development of heuristic methods for optimisation problems over rugged landscapes, where particles with memory could be used to complement gradient methods. 
Future work would be needed to ascertain the significance of memory for optimisation in high dimensions spaces and its effect in enhancing the efficiency of visual search.

\section*{Code availability}
The code to carry out the simulations and analysis can be found at \href{https://github.com/barahona-research-group/Chemotaxis-In-Rugged-Landscapes}{github.com/barahona-research-group/Chemotaxis-In-Rugged-Landscapes}  under DOI: \href{https://zenodo.org/badge/latestdoi/200467191}{10.5281/zenodo.3365951}. 

\section*{Data availability}

The data generated during the simulations is available with DOI: \href{https://doi.org/10.14469/hpc/6522}{10.14469/hpc/6522}.

\section*{Acknowledgements}
We thank Philipp Thomas, Jos\'{e} A.\ Carrillo and Julius B.\ Kirkegaard for insightful comments. We thank Eduardo Sontag, on one hand, and Nils Becker and Pieter Rein ten Wolde, on the other, for providing us with their agent-based codes, which helped the development and validation of our models. AG acknowledges funding through a PhD studentship under the BBSRC DTP at Imperial College (BB/M011178/1). MB acknowledges funding from the EPSRC project EP/N014529/1 supporting the EPSRC Centre for Mathematics of Precision Healthcare. \\

\section*{Author contributions}
AG and MB designed the study. AG performed the numerical and analytical work interacting closely with MB. AG and MB wrote the manuscript.

\section*{Competing interests}
The authors declare no competing interests.


\end{document}


\renewcommand\thesubsection{\arabic{subsection}}
\renewcommand{\figurename}{Supplementary Figure}
\renewcommand\theequation{S\arabic{equation}}

\begin{figure}
	\centering
	\includegraphics[width=0.8\textwidth]{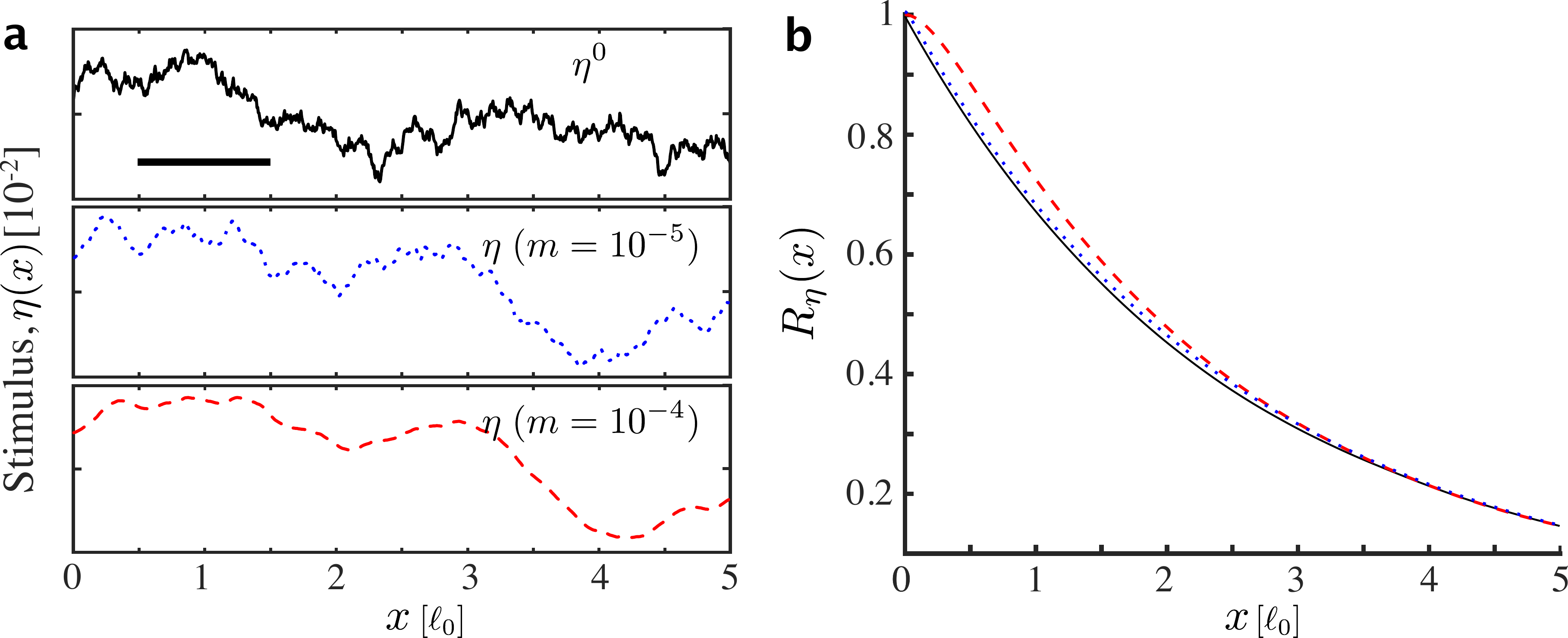}
	\caption{\textbf{Sample realisations and autocorrelation function of the harmonic oscillator $\eta(x)$ illustrating the deviation from a the limiting Ornstein-Ohlenbeck process $ \eta^0(x) $.} 
	\textbf{a} Sample trajectories of $ \eta^0(x) $ and $ \eta(x) $ for correlation length $ \mu=2 $ and regularisation factors $ m=10^{-4},\,10^{-5} $. Bar shows one run length. 
	\textbf{b} Autocorrelation function $ R_\eta(x):=\langle \eta(x)\eta(x') \rangle_\xi/\sigma^2_\eta $ corresponding to the trajectories in a. The departure from the OU process is not visible for $ m<10^{-5} $. We use the regularised process with $ m=10^{-5} $ for the computation of the agent-based and Keller-Segel models, while performing the analytical computations in the Ornstein-Uhlenbeck limit.  The trajectories for $ \eta(x) $ were obtained numerically by the Euler-Maruyama scheme with step $ \Delta x = 5\times 10^{-5} $. }
	\label{FigS5}
\end{figure}

\begin{figure}
	\centering
	\includegraphics[width=0.7\textwidth]{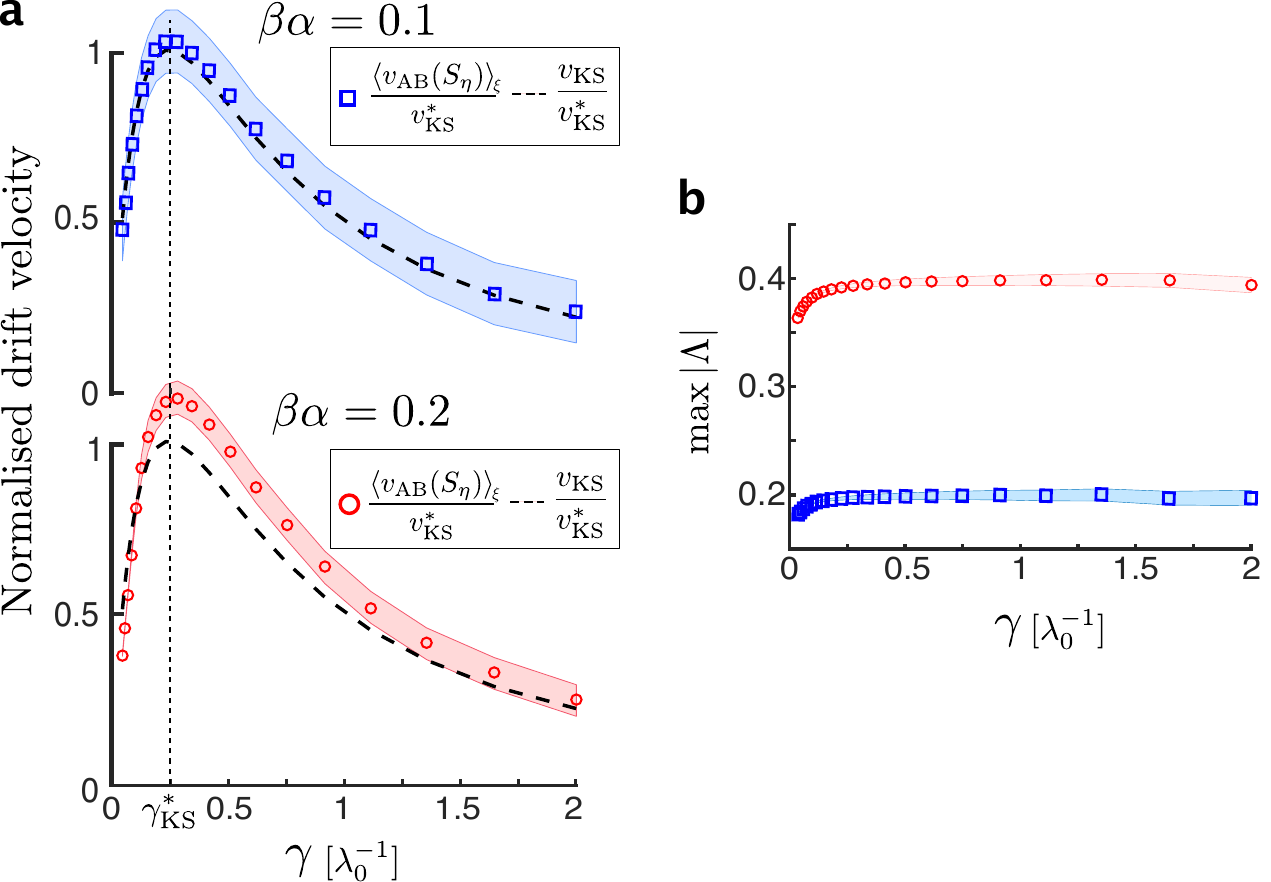}
	\caption{\textbf{Comparison of agent-based numerics and Keller-Segel approximation in shallow and steep gradients.} 
	\textbf{a} For small values of $\beta \alpha$ the Keller-Segel model accurately predicts the drift-velocity of agent-based cells for a large range of memories $\gamma$. For $\beta \alpha>0.1$ the average velocity of the agents is less well predicted by the Keller-Segel drift velocity. 
	\textbf{b} The reason for the discrepancy in a is that the small-response condition $\Lambda \ll 1$ no longer holds for large values of $\beta\alpha$ for all but very small memories $\gamma$.}
	\label{FigS6}
\end{figure}

\clearpage

\subsection{Supplementary Note}

\paragraph{Tumbling response}

\begin{figure}
	\centering
	\includegraphics[width=\textwidth]{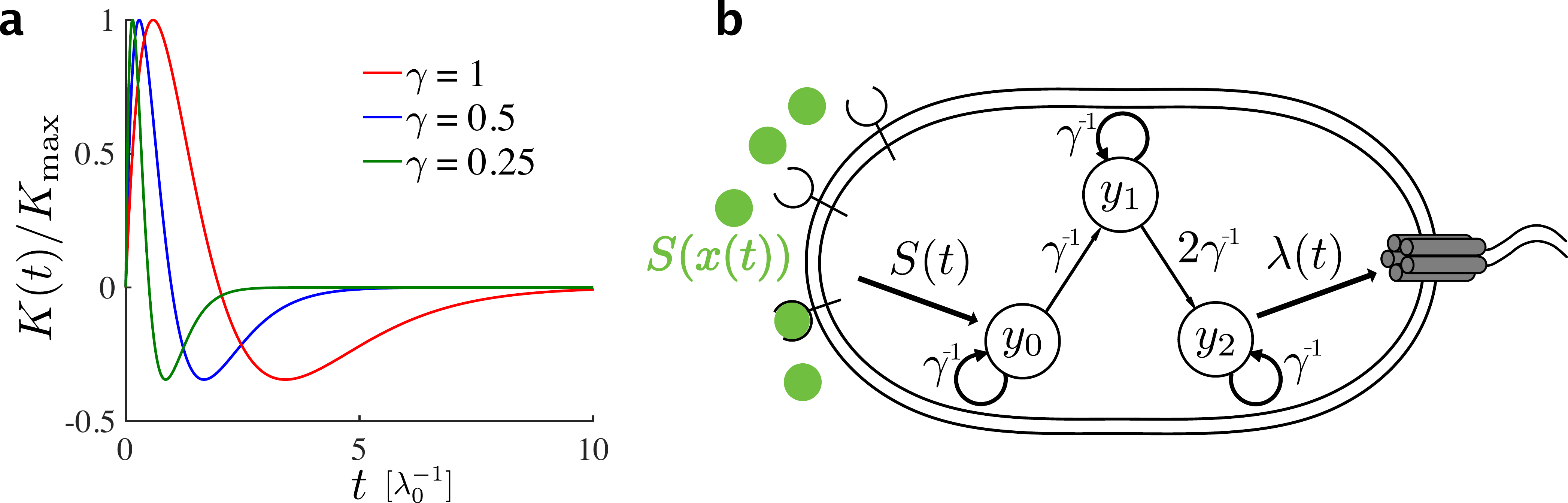}
	\caption{\textbf{Schematic of the 3-state model of cellular memory and impulse response dynamics.} 
	\textbf{a} The normalised response kernel at different values of the memory $ \gamma $ and as a function of time 
	\textbf{b} Schematic of the 3-state model of cellular memory, whose impulse response is $K(t) $ in a. The extracellular input signal $ S(x(t)) $ is transduced into a ligand concentration $ S(t) $, which modulates the activity of the 'kinase' $ y_0 $. This kinase relays the signal via an internal `buffer variable' $y_1$, which in turn produces the response regulator species $y_2$, akin to the CheY protein in the \textit{E. coli} system~\cite{Berg:2004}, in order to control the tumbling rate $ \lambda(t) $.}
	\label{FigS1}
\end{figure}

Consider cells adapted to the ambient concentration $ S(x_0) $ at the beginning of their trajectory $ x_0 $. The tumbling rate $ \lambda(t) $ at time $ t $ may then be modelled~\cite{tu,deGennes} by a linear perturbation depending on the chemoattractant concentrations $ S(x(u)) $ sensed along the trajectory $ x(u) $ in the recent past, $u<t $. Formally,
\begin{equation}
	\lambda (t)=1-\Lambda(t) \label{tmblrate}
\end{equation}
with the \textit{response} $ \Lambda(t) $ given by the convolution
\begin{equation}
\Lambda(t)=\int_{-\infty}^t K(t-u) S(x(u)) \, du.\label{response}
\end{equation}
Here, $ K(t) $ is the chemotactic kernel or response function, which for \textit{E. coli} adapting on aspartate has been experimentally characterised \cite{berg:74} as shown on Supplementary Figure \ref{FigS1}a. The bi-lobed shape gives positive weighting to recent observations and negative weighting to those further in the past with an exponential decay. As in \cite{clark,celani}, we consider the dimensionless scaling
\begin{equation}\label{Kt}
\begin{aligned}
	K(t)= 
	\begin{cases}
	\frac{\beta}{\gamma} \, e^{- t/\gamma} \left(\frac{t}{\gamma} - \frac{t^2}{2\gamma^2} \right) , &t\ge 0\\
	0, &t<0
	\end{cases}
\end{aligned}
\end{equation}
where, $\gamma$ is the relaxation time (i.e., the cellular \textit{memory}), $ \beta $ is the dimensionless signal gain, such that $ \beta/\gamma $ dictates the amplitude of the response kernel: $K_\text{max} = (\sqrt{2}-1) e^{\sqrt{2}-2} \beta/\gamma$.  

The empirical kernel $ K(t) $ has several important properties.
\begin{enumerate}[topsep=0pt,itemsep=-1ex,partopsep=1ex,parsep=0ex]
\item It is adaptive, that is $ \int K(t) d t = 0 $. \label{prop1} 
\item It is causal, i.e. $ K(t)=0 $ for $ t<0 $. Thus, one may exchange limits in the integral Eq. \eqref{response} $$ \Lambda(t)=\int_{-\infty}^t K(t-u) S(x(u)) \, du =  \int_0^\infty K(u) S(x(t-u)) \, du.$$ \label{prop2}
\item The kernel computes the instantaneous derivative of the perceived signal $ S(t) $ in the limit of vanishing memory $ \gamma\rightarrow 0 $. Indeed, using property \ref{prop2}, we have
\begin{align*}
\int_0^\infty K(u)S(t-u)d u & = \int_0^\infty K(u)(S(t)-u\frac{dS(t)}{dt}+\mathcal{O}( u^2))d u \\
& = \int_0^\infty K( \gamma w)( S(t)-\gamma w\frac{dS(t)}{dt}+\mathcal{O}(\gamma w^2))d w \\
& = \frac{dS(t)}{dt} \int_0^\infty |\gamma K(\gamma w)| dw-\mathcal{O}\left(\gamma\int_0^\infty K(\gamma w)w^2 dw\right) \\
& = \beta\frac{dS(t)}{dt} -\mathcal{O}\left(\gamma\int_0^\infty K(\gamma w)w^2 dw\right),
\end{align*}
where in the third equality we used property \ref{prop1}. Taking the limit $ \gamma \to 0 $ we obtain the result. \label{prop3}
\end{enumerate}

\paragraph{Internal dynamics}

The response Eq. \eqref{response} can be obtained from a linear dynamical system with three internal variables $ \bm{y} =(y_0,y_1,y_2) $, as shown by \cite{celani}. Let 
\begin{equation}\label{output}
	\Lambda(\bm{y}(t))= \beta y_2
\end{equation}
be the variable modulating the output. Then, using Eq. \eqref{response},
\begin{align}\label{y_2}
	\frac{d}{dt}y_2=\frac{d}{dt}\int_{-\infty}^t\frac{1}{\gamma} e^{-(t-u)/\gamma}\left[\frac{1}{\gamma}(t-u)-\frac{1}{2\gamma^2}(t-u)^2\right]S(x(u))\,du\,,
\end{align}
and we may take account of the memory by enlarging the state space to include a finite number of internal degrees of freedom transforming the infinite dimensional integro-differential equation Eq. \eqref{y_2} into a finite dimensional system of ODEs by the linear chain trick~\cite{macdonald}. Repeatedly differentiating Eq. \eqref{y_2} by Leibniz's rule we obtain 
\begin{equation}\label{intstates}
	\frac{d\bm{y}}{dt} = \bm{Ay} + \bm{b}\,,
\end{equation}  
where
\begin{equation*}
\bm{A}= \left(\begin{matrix} -1/\gamma &0&0\\ 1/\gamma &-1/\gamma &0\\0& 2/\gamma & -1/\gamma \end{matrix}\right), \quad\bm{b} = \left(\begin{matrix} \nabla S(x(t))\cdot v \\0\\0\end{matrix}\right).
\end{equation*}
In this dynamical system, $ y_0 $ is the input node sensing the changes $S(x(t))$ perceived by a moving cell; $ y_2 $ is the output node controlling the propensity of tumbling; and $ y_1 $ is the regulatory node, which acts as a 'buffer' integrating the difference between network response and steady-state output (Supplementary Figure \ref{FigS1}a. The unique adapted state (steady-state) of Eq. \eqref{intstates} is $ \bm{y}_\infty=(0,\,0,\,0) $, and the eigenvalues of the Jacobian matrix $ \bm{A} $ are negative $ (-1/\gamma,\,-1/\gamma,\,-1/\gamma) $, such that $ \bm{y}_\infty $ is locally asymptotically stable. In the following, we will also assume that $ \bm{y}(t) $ remains bounded.

\paragraph{Agent-based (AB) model}

We introduce the agent-based (AB) modelling framework based on Refs.~\cite{rousset,becker}. At the microscopic level the evolution of the state of a single cell, $ t\mapsto (x,v,\bm{y}) $, can be described as 
\begin{equation}\label{ABmod}
\begin{dcases}
	&\frac{dx}{dt} = v(t)\,,\quad \textit{(position)}\\ 
	&\frac{d\bm{y}}{dt} = \bm{Ay} + \bm{b}\,,\quad \textit{(internal state)}\\
	&\int_{t_{n-1}}^{t_n} \lambda(\bm{y}(t))dt =2\psi_n\quad \textit{(jump times)} \\
	&v(t) = \nu_n,\quad t\in [t_{n-1},t_{n}), \quad \nu_n = -\nu_{n-1} \quad \textit{(velocity-jumps)}.
\end{dcases}
\end{equation}

\subsection{Supplementary Note}

\paragraph{Connection to other modelling frameworks for chemotaxis response}

Several other works~\cite{tu,dufour} describe the tumbling rate modulations based on the phosphorylation dynamics of the CheY response regulator protein. Although the full model takes account of the non-linear binding kinetics between receptor and chemoattractant molecules, it is typically linearised assuming that the attractant concentration saturates the active receptor population and does not bind to the inactive receptor population. Under such conditions~\cite{dufour}, the tumbling rate is given by: 
$$\lambda(t)=\lambda_0 -\beta F(t), $$ 
with the internal state $F$ evolving according to
\begin{equation}\label{internalstate}
    \frac{dF}{dt} = -\frac{1}{\gamma}(F-F_0) + \omega S'(x),
\end{equation}
where $\tau$ is the adaptation timescale (memory); $S'(x)$ is the chemoattractant gradient; and the constant $\omega = \pm\nu N$ is the product of the running velocity $\nu$ and the receptor gain $N$.

It is easy to show that the tumbling rate can then be equivalently expressed in the form of Eqs. \eqref{tmblrate}--\eqref{response}. 
Indeed, integrating Eq. \eqref{internalstate} by parts, one obtains
\begin{align*}
    F &= F_0 + \omega e^{-t/\gamma}\left( \left[ e^{u/\gamma}S(u) \right]_0^t - \frac{1}{\gamma}\int_0^t e^{u/\gamma}S(u)\,du \right) \\
    &= F_0 + \omega  e^{-t/\gamma}S(0) + \omega \int_0^t \left( \delta(t-u) - \frac{1}{\gamma} e^{-(t-u)/\gamma} \right)S(u)\,du \\
    &= \omega \int_0^t K(t-u)S(u)\,du,
\end{align*}
assuming that initially $S(0) = 0$ and the adapted state $F_0=0$. This expression is in the required form with 
\begin{align}
K(t)=\delta(t) - (1/\gamma) e^{-t/\gamma},
\label{eq:dufour_kernel}
\end{align}
where $\delta(t)$ is the Dirac-delta function.
Note that the kernel Eq. \eqref{eq:dufour_kernel} consists of a singularity at time $t=0$ combined with a single decaying exponential.  In contrast, the kernel we consider in our work starts at the origin and is characterised by a finite excitation time and finite adaptation time leading to one negative overshoot and one positive overshoot.

\subsection{Supplementary Note}

\paragraph{Derivation of the Keller-Segel drift velocity based on local gradient alignment}\label{sect:ksdrift}
 
In this section, we follow the derivation of the Keller-Segel drift velocity by deGennes \cite{deGennes}. However, instead of a constant gradient, we consider a constant gradient with additive noise, modelled by $ S_\eta $. In the rest of this section, we work in the OU limit ($m\to 0$).

Drift velocity is generated when the average runs up the gradients are longer than those the way down. The dimensionless drift velocity on ballistic timescales is therefore
\begin{equation}
	v_\text{AB}(S_\eta)\simeq  \frac{\langle t^+\rangle_\text{AB} - \langle t^- \rangle_\text{AB}}{\langle t^+\rangle_\text{AB} + \langle t^- \rangle_\text{AB}} = \langle t^+\rangle_\text{AB} - 1,\tag{18}
\end{equation}
where $  t ^+ $ and $  t ^- $ are average run times over all possible trajectories for cells moving up and down the gradient for a given realisation of $ S_\eta $. As described in the main text, the drift velocity averaged over realisations of the landscape $S_\eta$ becomes
\begin{equation}
    \vABnoise \simeq \int_0^\infty e^{-s} \left\langle \int_0^s\Lambda(t)dt\right\rangle_\xi ds + \frac{1}{2} \int_0^\infty e^{-s} \left \langle \left(\int_0^s \Lambda(t)dt \right)^2 \right\rangle_\xi ds. \tag{20}
\end{equation}

\paragraph*{Remark} 
Although $|\Lambda(t)|\ll 1$ suffices to justify the expansion Eq. (20), note that it also holds when $ \int_0^s \Lambda(t)dt \ll 1$. This is a weaker assumption since $ \Lambda(t) $ can be negative causing cancellations to occur in the integral. Also, as argued in the main text, $|\Lambda(t)|\ll 1$ is less restrictive than the classical shallow-gradient assumption of Erban et al. \cite{erban}, since it takes account of the internal timescale $ \gamma $ in relation to the perceived signal. 

In section we describe how to evaluate the first term in Eq. (20), following \cite{deGennes}, while leaving the second term for Supplementary Note \ref{sect:computationof}. Throughout, we will assume that the small response condition holds. In Supplementary Note \ref{momentclos} we derive conditions on $ \alpha,\;\beta,\;\sigma_\eta $ that guarantees that it holds, both in constant gradient and in the presence of additive noise. 

Then, to obtain the first term, we first compute
\begin{align}
\left\langle\int_0^s\Lambda(t) dt \right\rangle_\xi =& \int_0^s \int_0^\infty K(u)\langle S_\eta(t-u)\rangle_\xi du dt = \int_0^s \int_0^\infty K(u) S_0(x(t-u)) du dt\,. \tag{21}\label{firstterm}
\end{align}
Then, following de Gennes, we use techniques from linear response theory; first, due to the linearity of the response $ \Lambda(t) $ we use linear superposition $ K(u) = \int_0^\infty K(t)\delta(t-u)du $, and second, assuming that the gradient does not change over a run, we expand $ S_0(t-u) = S_0(0) + \frac{dS_0(0)}{dx}(t-u) = S_0(0) + \alpha\,(t-u) $. Then, we have
\begin{align*}
\left\langle\int_0^s\Lambda(t) dt \right\rangle_\xi =& \int_0^\infty K(u)\int_0^s \left(S_0(0) + \alpha(t-u)\right) dtdu = \alpha\int_0^\infty K(u)\int_0^s(t-u) dt du\,,
\end{align*}
where in the first equality we used Property \ref{prop1} of $K(t)$. Further, assuming that the positions before and after the tumble at time $ t_n = 0$ are not correlated, one may truncate the kernel $ K(t) $ at times $ t<u $. Therefore,
\begin{align}
\left\langle\int_0^s\Lambda(t) dt \right\rangle_\xi \simeq & \, \alpha \int_0^\infty K(u)\int_u^s(t-u) dt du\notag = \frac{\alpha }{2}\int_0^\infty K(u)(s-u)^2 du\,. \label{t1}
\end{align}
Inserting into (20) and integrating over $ s $ and $ u $ yields de Gennes's result \cite{deGennes} for constant gradients
\begin{align}
\vABnoise \simeq \vKS &= \frac{\alpha}{2}\int_0^\infty K(u)\left(\int_u^\infty e^{-s} (s-u)^2 ds \right) du
= \frac{2\beta\alpha\gamma}{(1+2\gamma)^3}\,. \tag{24}
\end{align}
Note that no contribution from the spatial randomness appears, since these cancel upon taking averages. 

\subsection{Supplementary Note}\label{momentclos}

\paragraph{Moment closure}

The above derivation of the drift velocity Eq. \eqref{vKScorr} relies on the small response assumption
\begin{equation}
|\Lambda(t)| \ll 1, \quad{(\textit{small-response})}.\tag{7}
\end{equation}
We now turn to computing the parameter regimes for which this assumption holds, first for constant gradients and then for gradients with additive noise.
   
\paragraph{Small-response condition in constant gradients}

In constant gradients, the small response condition can be explicitly computed. We compute the output variation over one run:
\begin{align*}
	\Lambda(t)&=\int_0^\infty K(u) S_0(t-u) \, du 
	= \frac{\beta\alpha}{\gamma}\int_0^\infty  \, e^{- u/\gamma} \left(\frac{u}{\gamma} - \frac{u^2}{2\gamma^2} \right) (t-u) \, du =\beta\alpha\gamma ,
\end{align*}
where we used integration by parts. Hence the amplitude is $ |\Lambda(t)| = \beta\alpha\gamma  $, and for memories up to $\gamma \sim 1 $, we have that the condition
\begin{equation}
\beta\alpha \ll 1\quad \textit{(shallow perceived gradient)}\label{smallab}
\end{equation}
is sufficient to ensure that Eq. (7) holds in the main text, i.e., that the cells perceive small variations of the chemoattractant.

\paragraph{Small-response condition in constant gradients with additive noise}

Unlike in the constant gradient case, there is no closed form expression for the response amplitude $ |\Lambda| $ when the cell is moving in a random landscape $ S_\eta $. Therefore we provide an upper bound as a function of the internal scales $ \beta,\, \gamma $ and the external stimulus $ \sigma_\eta,\, \mu $. The results below assume that the noise on the landscape is additive, but not otherwise independent of the stochastic model chosen. The following relies on a concentration argument by bounding the mean and the variance of the response amplitude, $\langle |\Lambda| \rangle_\xi $ and $Var_\xi(|\Lambda|) $, respectively, where the averages $ \langle \cdot \rangle_\xi $ are taken over the distribution of $ \eta^0 $. 

We first find an upper bound on the average response, $ \langle|\Lambda|\rangle_\xi $ in terms of the constants $ \alpha,\,\beta,\,\sigma_\eta^2 $. We use the scaled kernel $ \widetilde{K}(t) $. Then, by the Cauchy-Schwarz inequality we have:
\begin{align}
\langle |\Lambda(t)| \rangle_\xi^2 \le \langle |\Lambda(t)|^2 \rangle_\xi &=  \int_0^\infty\int_0^\infty \widetilde{K}(u)\langle S_{\eta^0}(t-u)S_{\eta^0}(t-v)\rangle_\xi\widetilde{K}(v) dudv \notag \\
&= \left(\int_0^\infty\widetilde{K}(u)\alpha(t-u)du\right)^2 + \int_0^\infty\int_0^\infty \widetilde{K}(u)\langle\eta^0(t-u)\eta^0(t-w)\rangle_\xi\widetilde{K}(w) dudw \notag \\
&=\beta^2(\alpha^2 + \sigma^2_\eta\sigma_{\Lambda}^2)\,,
\end{align}
where $ \sigma_{\Lambda}^2 $ is the normalised response variance.  

\begin{figure}[bt]
	\centering
	\includegraphics[width=0.5\textwidth]{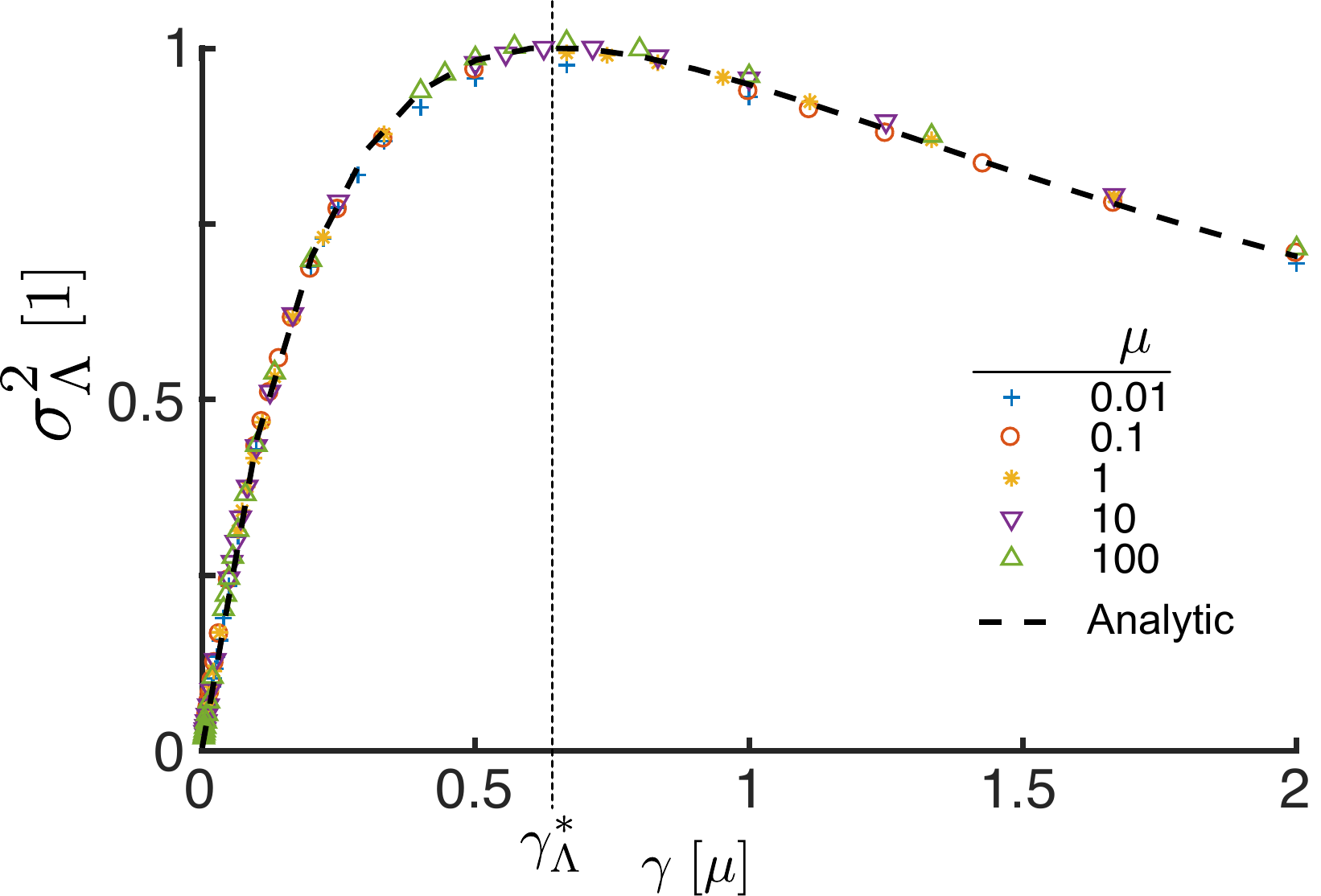}
	\caption{\textbf{Tumbling rate variance $\sigma^2_\eta$ as a function of $\Gamma=\gamma/\mu$, the memory length $\gamma$ relative to the correlation length $\mu$ of the Ornstein-Uhlenbeck input $\eta^0$}. Symbols show Monte Carlo simulations using the linear dynamical system Eq. \eqref{intstates} driven by $\eta^0$ at various values of $\mu$. Dashed line shows analytical computations using Eq. \eqref{sigmalambda}. The maximum variance is attained at $ \gamma^*_\Lambda = (\sqrt{7}-2)\mu $. The trajectories for $ \eta^0 $ were obtained numerically using the Euler-Maruyama scheme with step $ \Delta x = 10^{-4} $.}  
	\label{FigS2}
\end{figure}

In the specific case, when the input is a stationary  OU process $ \eta^0 $, the latter can be explicitly computed, 
\begin{equation}
\sigma^2_{\Lambda}(\Gamma) = \frac{\Gamma(\Gamma + 3)}{8(\Gamma+1)^3}\,, \label{sigmalambda}
\end{equation}
where $\Gamma = \gamma/\mu$, defining the ratio between memory and perceived correlations. The stationary points with respect to $ \Gamma $ yield a unique maximum $ \gamma_\Lambda^* = (\sqrt{7}-2)\mu$. Eq. \eqref{sigmalambda} is plotted on Supplementary Figure \ref{FigS2}, showing excellent agreement between theory and simulations.

Next, we compute an upper bound on the variance $ Var(|\Lambda|) $. Note that 

\begin{align*}
Var_\xi(|\Lambda|) &= \langle |\Lambda|^2 \rangle_\xi - \langle |\Lambda| \rangle_\xi^2 =\langle \Lambda^2 \rangle_\xi - \langle |\Lambda| \rangle_\xi^2 =Var_\xi(\Lambda) + \langle \Lambda \rangle_\xi^2 - \langle |\Lambda| \rangle_\xi^2 \le Var_\xi(\Lambda)\,,
\end{align*}
since $ \langle \Lambda \rangle_\xi \le \langle |\Lambda| \rangle_\xi$. Therefore, it suffices to find a bound on the variance of $ \Lambda $. We have
\begin{align}
Var_\xi(\Lambda) &= \langle\Lambda^2\rangle_\xi -  \langle\Lambda\rangle_\xi \notag\\
&=\left \langle \int_0^\infty\int_0^\infty K(u)S_\eta(t-u)S_\eta(t-w)\widetilde{K}(w)\, du dw\right\rangle_\xi -\left(\left \langle \int_0^\infty K(u)S_\eta(t-u) du\right\rangle_\xi\right)^2\notag\\
&=\beta^2\alpha^2 +  \int_0^\infty\int_0^\infty K(u)\langle\eta^0(u)\eta^0(w)\rangle_\xi K(w) \, du dw - \beta^2\alpha^2 =\beta^2\sigma_\eta^2\sigma^2_{\Lambda}\gamma^2\,,
\end{align}
where we used the stationarity of $ \eta^0 $ and the definition of $\sigma_\Lambda$ in Eq. \eqref{sigmalambda}.

Collating the above results, we have the following bounds on the mean and variance
\begin{align}
&\langle |\Lambda| \rangle_\xi \le \beta\sqrt{\alpha^2 + \sigma^2_\eta\sigma_\Lambda^2}\label{responsemean}\\
& Var_\xi(|\Lambda|)\le \beta^2\sigma^2_\eta\sigma_\Lambda^2 \,. \label{responsevar}
\end{align}

Finally, we use the bounds Eqs. \eqref{responsemean}--\eqref{responsevar} to control the probability of large response amplitudes. By Chebyshev's inequality we obtain 
\begin{equation}
\mathbf{P}\left(|\Lambda - \langle\Lambda\rangle_\xi|\ge \epsilon \right) \le \frac{\beta^2 \sigma_\eta^2\sigma_\Lambda^2}{\epsilon^2} \le \frac{c\beta^2 \sigma_\eta^2}{\epsilon^2} \,,\label{cheby}
\end{equation}
where $ c=(10+7\sqrt{7})/54 $ is the maximum of $ \sigma_\Lambda^2 $ in Eq. \eqref{sigmalambda} independently of the system parameters. In the constant gradient limit, as $\mu\to\infty$ or $ \sigma_\eta\to 0 $, we have $ \sigma_\Lambda^2 \to 0 $ and hence $ P (|\Lambda-\langle\Lambda\rangle_\xi|\ge \epsilon) \to 0 $. Thus, $\beta\alpha\ll 1$ is sufficient to control the response, recovering Eq. \eqref{smallab}. Likewise, in the white noise limit, as $\mu\to 0$, we also have $ \sigma_\Lambda^2 \to 0 $. Therefore, we see that no additional condition on $ \sigma^2_\eta $ is necessary for the KS equation to be valid. 

However, in general, when the correlation length is not restricted, i.e. $ 0<\mu<\infty $, the condition $\beta\alpha\ll 1$ is no longer sufficient for the small response condition to hold, but due to Eq. \eqref{responsemean} and Eq. \eqref{cheby} additional conditions are needed: $\beta\sqrt{\alpha^2+c\sigma^2_\eta}= \mathcal{O}(\epsilon)$ and $\sqrt{c}\beta\sigma_\eta \ll \mathcal{O}(\epsilon)$. These are equivalent to requiring
\begin{equation}
\begin{aligned}\label{smallcoeff}
&\beta\alpha \ll 1\quad &&\textit{(shallow constant gradient)} \\
&\frac{\alpha}{\sigma_\eta} \gg 1 &&\textit{(large signal-to-noise ratio)},
\end{aligned}
\end{equation}
which means that in addition to Eq. \eqref{smallab}, we require signal-to-noise ratio to be large. 

\subsection{Supplementary Note}\label{sect:computationof}

\paragraph{Computation of the second-order contribution}

\begin{figure}
	\centering
	\includegraphics[width=.9\textwidth]{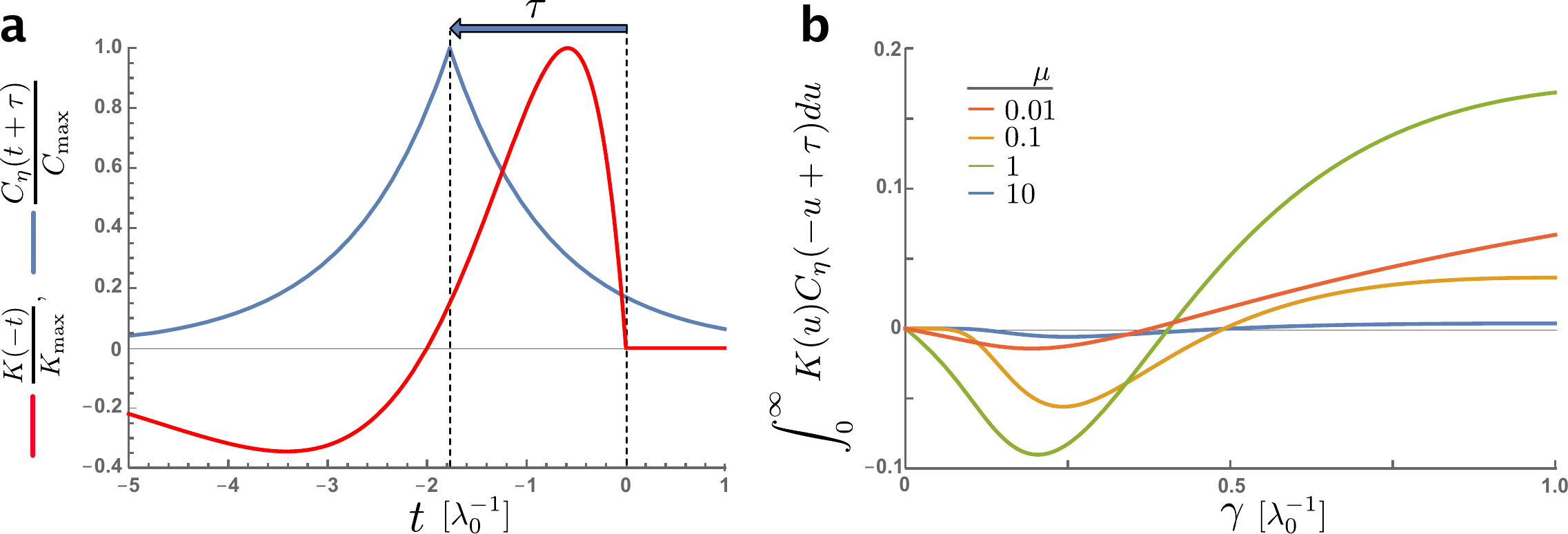}
	\caption{\textbf{Relationship between the response kernel and the correlation structure of the environment, for finite response time}. \textbf{a} Causal responses have a non-negative delay between the input and the output $\tau\ge 0 $, which is the case in chemotaxis due to the finite time between sensing the tumbling taking place. This delay results in a time-shift between the response kernel $K(t)$ and the delayed autocorrelation function $C_\eta(t+\tau)$. \textbf{b} As a result, for a fixed $\tau$, small memories $\gamma$ lead to suboptimal filtering. Plotting the overlap integral between the memory kernel and the delayed autocorrelation function as a function $\gamma$ and for different values of the correlation length $\mu$ shows that for a given $\mu$ small $\gamma$ yields a negative contribution to the overlap integral and hence the drift speed (Eq. \eqref{l1l1}).}
	\label{FigS3}
\end{figure}

To obtain the contribution from the noise, we begin with 
\begin{equation}
    \vABnoise \simeq \int_0^\infty e^{-s} \left\langle \int_0^s\Lambda(t)dt\right\rangle_\xi ds + \frac{1}{2} \int_0^\infty e^{-s} \left \langle \left(\int_0^s \Lambda(t)dt \right)^2 \right\rangle_\xi ds. \tag{20}
\end{equation}
At this point, we depart from de Gennes' derivation and compute the second-order term to obtain the contribution from the rugged landscape. We proceed by rearranging the integrals
\begin{align}
\left\langle\left(\int_0^s\Lambda(t)dt\right)^2\right\rangle_\xi = \left\langle\int_0^s\int_0^s\Lambda(t)\Lambda(\tilde{t})dtd\tilde{t}\right\rangle_\xi = \int_0^s\int_0^s\left\langle\Lambda(t)\Lambda(\tilde{t})\right\rangle_\xi dtd\tilde{t}\,. \label{term2}
\end{align}
Then, using the definition $ S_\eta:= S_0 + \eta^0 $, we consider the integrand
\begin{align}
\langle\Lambda(t)\Lambda(\tilde{t})\rangle =&  \int_0^\infty\int_0^\infty K(w)S_{\eta}(x(\tilde{t}-w))\langle S_{\eta}(x(t-u))\rangle_\xi K(u) du dw\notag\\
=& \int_0^\infty\int_0^\infty K(w)S_0(x(\tilde{t}-w))S_0(x(t-u))K(u)\, du dw \notag \\
& \quad 
+ \int_0^\infty\int_0^\infty K(w) \langle\eta^0(x(\tilde{t}-w))\eta^0(x(t-u))\rangle_\xi K(u)\,du dw \notag\\
=& \int_0^\infty\int_0^\infty K(w)S_0(x(t-u))S_0(x(\tilde{t}-w))K(u) \,du dw \notag \\
& \quad
+ \int_0^\infty K(w) \left( \int_0^\infty K(u) C_\eta(x(\tau-u))\,du \right) dw\,, \label{l1l1}
\end{align}
where the term in brackets defines the overlap integral between the autocovariance function of the input and the chemotactic kernel shifted by $\tau = t-\tilde{t}+w$. 
This integral is depicted on Supplementary Figure \ref{FigS3} has no contribution in the limits $\mu\to 0,\,\infty$ since the autocovariance $C_\eta$ is approximately constant and by property \ref{prop1} of $K(t)$. Between these limits, depending on $\mu$, there is a negative contribution for small memories signifying suboptimal filtering and positive contribution for large memories.

Combining Eqs. \eqref{term2}--\eqref{l1l1}, we have
\begin{align}
\left\langle\left(\int_0^s\Lambda(t)dt\right)^2\right\rangle_\xi&= \left(\int_0^s\int_0^\infty  K(u)S_0(t-u) du dt \right)^2 \notag\\
&\quad + \int_0^\infty K(w)\left( \int_0^s\int_0^s C_{\eta\Lambda}(t-u-\tilde{t}+w)  \,d\tilde{t} dt  \right)dw\\
&= \left(\int_0^s\int_0^\infty  K(u)S_0(t-u) du dt \right)^2 \notag\\
&\quad + \int_0^\infty K(w)\left( \int_w^s\int_0^{\tilde{t}-w} C_{\eta\Lambda}(t-u-\tilde{t}+w)  \,d\tilde{t} dt  \right)dw\,. \label{term22}
\end{align}
Note, the first term on the right hand side is the square of Eq. \eqref{firstterm}, and therefore it is treated as in Supplementary Note \ref{sect:ksdrift} to yield $\vKS^2 $. In the second term, we first exchanged the integrals, which is allowed since the integrand is bounded due to the exponential decay of $ K(t) $, Eq. \eqref{Kt}. Second, as in deGennes' derivation, we ignored the contribution of inputs before the tumbling at time $ t=0 $, i.e., $  t\in[w,s] $. Thus, information obtained before the tumble does not contribute to the drift speed. Third, due to causality, we need $\tau = t-\tilde{t}+w \ge 0$, i.e., the input at $\eta^0(t-\tau) $ must preceed the output at $\Lambda(t)$. As a result, we only obtain contribution in the region $\tilde{t}\times t \in [w,s]\times [0,\tilde{t} - w]$, as shown by the hashed area in Supplementary Figure \ref{FigS4}.

\begin{figure}
	\centering
	\includegraphics[width=0.35\textwidth]{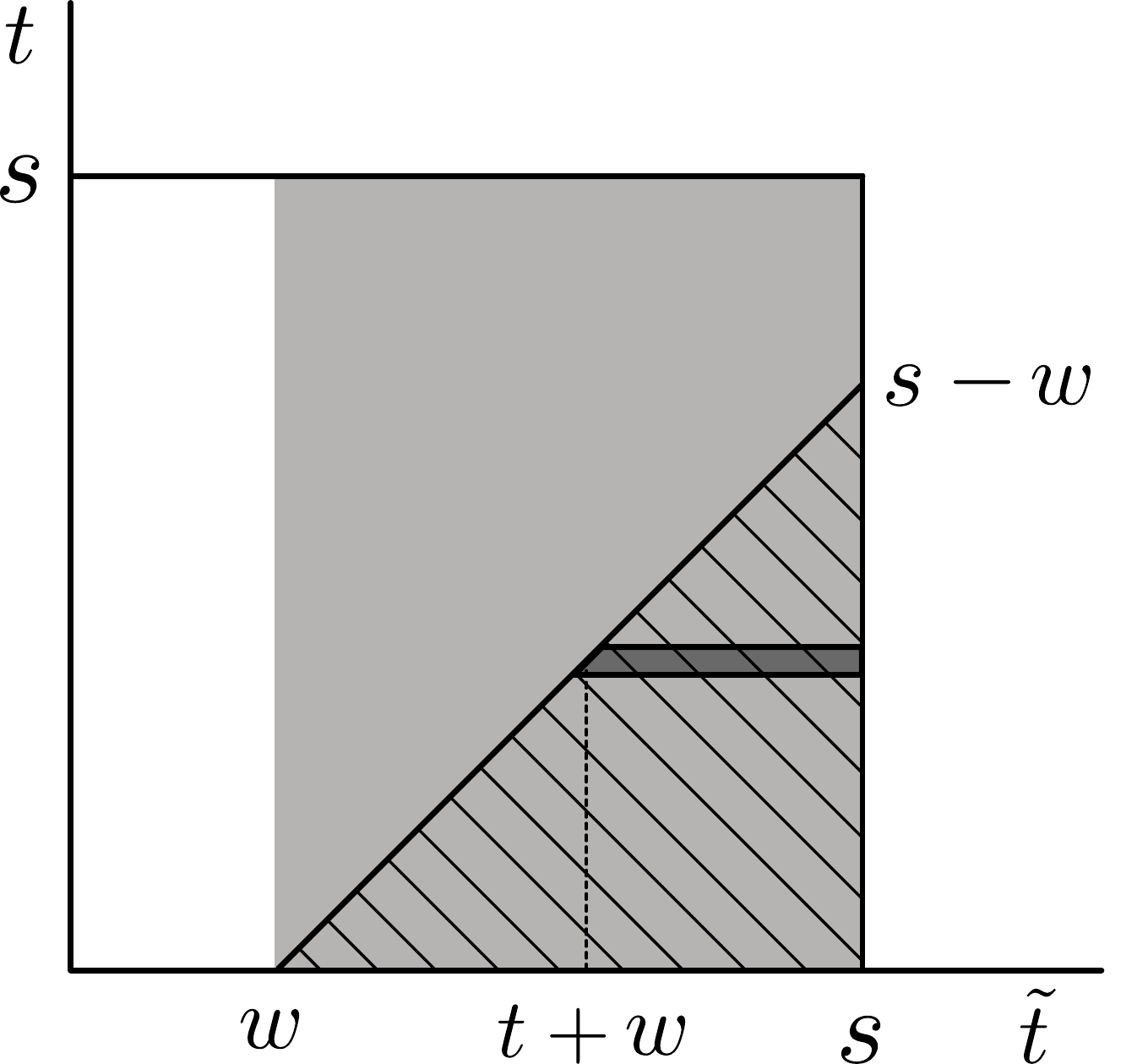} \vspace*{-.01in}
	\caption{\textbf{Schematic illustrating the integration region in Eq. \eqref{term22} that is consistent with causal responses.} The x and y-axes show the two variables of integration $t,\;\tilde{t}$. Since the swimming direction before and after a tumble are uncorrelated, inputs registered before the tumble (at $t=0$) do not contribute to the integral. The grey area represents the region of integration to the integral after the tumble. Second, due to causality, the input $\eta^0(\tilde{t}-w)$ must precede the output $\Lambda(t)$. This corresponding integration area is given by the hashed region. Finally, for convenience, we evaluate the double integral by scanning the dark grey area over the hashed region.}
	\label{FigS4}
\end{figure}

With these assumptions, the correction term due to the random OU fluctuations can be evaluated. We therefore let $\tau= t - \tilde{t}+w $ and consider the input-output covariance in Eq. \eqref{term22}
\begin{align}\label{iocovar}
C_{\eta\Lambda}\left(\tau\right) &= \langle \eta^0(t)\Lambda(t+\tau ) \rangle_\xi  = \int_0^\infty K(t)R_\eta(\tau ) dt \notag \\
&=\frac{\beta\sigma^2_\eta}{\gamma} \int_0^\infty \exp\left(-\frac{t}{\gamma}-\frac{|t-\tau|}{\mu}\right) \left(\frac{t}{\gamma}-\frac{t^2}{2\gamma^2}\right)dt \notag \\
&= \beta\sigma_\eta^2\begin{cases}
\; e^{-\frac{\tau }{\mu }}\frac{ \Gamma }{(\Gamma-1 )^3 } 
- e^{-\frac{\tau }{\gamma }} \biggl[ 
\frac{2 \Gamma \left(3 \Gamma ^2+1 \right)}{ (\Gamma^2-1 )^3 }
 - \frac{2\left(\Gamma^2+1\right)}{(\Gamma^2-1 )^2 } \frac{ \tau}{\mu} 
 + \frac{ 1}{(\Gamma^2-1 ) } \frac{\tau^2}{\gamma\mu}  
  \biggr] \,, \quad \text {if }\tau>0 \\
  \;\frac{\Gamma}{(\Gamma + 1)^3}, \quad\text{if } \tau=0
  \end{cases}
\end{align}
where $ \Gamma =\gamma/\mu $, as before. Note that although $ (\Gamma^2-1) $ appears in the denominator, no blow-up occurs when $ \Gamma = 1 $, as expected since the integrals converge. Combining Eqs. \eqref{term22} - \eqref{iocovar} and integrating the second term in Eq. \eqref{term22} may be evaluated to yield
\begin{align}\label{t2}
\left\langle\left(\int_0^s\Lambda(t)dt\right)^2\right\rangle_\xi = \left(\frac{2\beta\alpha\gamma}{(1+2\gamma)^3}\right)^2 + \beta^2 \sigma_\eta^2\frac{ \gamma ^2 \mu  \left[2 \gamma ^3 (1+\mu) + (1+\gamma)^3 \mu ^2  + 6 \gamma ^2 \mu  -2  \mu ^2\right]  }{(1+\gamma )^6 (1+\mu ) (\gamma +\mu )^3}
\end{align}
Finally, collating Eqs. (20), (24) in the main text and \eqref{t2}, we obtain the drift velocity constant landscapes with additive OU noise keeping only those terms at second order originate from the noise $\eta^0$
\begin{align}
\vABnoise := \Vmu& 
\simeq \frac{\alpha\beta \gamma}{(1+2\gamma)^3} + \frac{\beta^2 \sigma_\eta^2}{2}\frac{ \gamma ^2 \mu  \left[2 \gamma ^3 (1+\mu) + (1+\gamma)^3 \mu ^2  + 6 \gamma ^2 \mu  -2  \mu ^2\right]  }{(1+\gamma )^6 (1+\mu ) (\gamma +\mu )^3} =\vKS + \Delta v_\mu \,. \label{vKScorr}
\end{align}

\clearpage
{\normalfont\Large\bfseries \noindent Supplementary References}